\documentclass[12pt]{article}
\usepackage{graphicx}

\def\hybrid{\topmargin 0pt      \oddsidemargin 0pt
        \headheight 0pt \headsep 0pt
       \voffset-1cm
        \textwidth 6.25in       
       \textheight 9.5in       
        \marginparwidth 0.0in
        \parskip 5pt plus 1pt   \jot = 1.5ex}
\catcode`\@=11
\def\marginnote#1{}

\newcount\hour
\newcount\minute
\newtoks\amorpm
\hour=\time\divide\hour by60
\minute=\time{\multiply\hour by60 \global\advance\minute by-\hour}
\edef\standardtime{{\ifnum\hour<12 \global\amorpm={am}%
        \else\global\amorpm={pm}\advance\hour by-12 \fi
        \ifnum\hour=0 \hour=12 \fi
        \number\hour:\ifnum\minute<10 0\fi\number\minute\the\amorpm}}
\edef\militarytime{\number\hour:\ifnum\minute<10 0\fi\number\minute}

\def\draftlabel#1{{\@bsphack\if@filesw {\let\thepage\relax
   \xdef\@gtempa{\write\@auxout{\string
      \newlabel{#1}{{\@currentlabel}{\thepage}}}}}\@gtempa
   \if@nobreak \ifvmode\nobreak\fi\fi\fi\@esphack}
        \gdef\@eqnlabel{#1}}
\def\@eqnlabel{}
\def\@vacuum{}
\def\draftmarginnote#1{\marginpar{\raggedright\scriptsize\tt#1}}

\def\draftlabel#1{{\@bsphack\if@filesw {\let\thepage\relax
   \xdef\@gtempa{\write\@auxout{\string
      \newlabel{#1}{{\@currentlabel}{\thepage}}}}}\@gtempa
   \if@nobreak \ifvmode\nobreak\fi\fi\fi\@esphack}
        \gdef\@eqnlabel{#1}}
\def\@eqnlabel{}
\def\@vacuum{}
\def\draftmarginnote#1{\marginpar{\raggedright\scriptsize\tt#1}}

\def\draft{\oddsidemargin -.5truein
        \def\@oddfoot{\sl preliminary draft \hfil
        \rm\thepage\hfil\sl\today\quad\militarytime}
        \let\@evenfoot\@oddfoot \overfullrule 3pt
        \let\label=\draftlabel
        \let\marginnote=\draftmarginnote
   \def\@eqnnum{(\theequation)\rlap{\kern\marginparsep\tt\@eqnlabel}%
\global\let\@eqnlabel\@vacuum}  }

\def\numberbysection{\@addtoreset{equation}{section}
        \def\theequation{\thesection.\arabic{equation}}}

\def\underline#1{\relax\ifmmode\@@underline#1\else
        $\@@underline{\hbox{#1}}$\relax\fi}

\def\titlepage{\@restonecolfalse\if@twocolumn\@restonecoltrue\onecolumn
     \else \newpage \fi \thispagestyle{empty}\c@page\z@
        \def\thefootnote{\fnsymbol{footnote}} }

\def\endtitlepage{\if@restonecol\twocolumn \else  \fi
        \def\thefootnote{\arabic{footnote}}
        \setcounter{footnote}{0}}  
\relax

\numberbysection
\hybrid

\newfont{\Bbb}{msbm10 scaled 1\@ptsize00}
\newfont{\Bbbb}{msbm7 scaled 1\@ptsize00}
\newcommand{\CC}{\mbox{\Bbb C}}

\newcommand{\DDD}{\raise-1pt\hbox{$\mbox{\Bbbb D}$}}

\newcommand{\UUU}{\raise-1pt\hbox{$\mbox{\Bbbb U}$}}

\newcommand{\ZZ}{\mbox{\Bbb Z}}
\newcommand{\z}{\raise-1pt\hbox{$\mbox{\Bbbb Z}$}}

\def\beq{\begin{equation}}
\def\eeq{\end{equation}}
\def\p{\partial}

\def\res{\mathop{\hbox{res}}\limits}

\newtheorem{lemma-definition}{Lemma-Definition}[section]

\begin{document}

\begin{titlepage}

\title{Quasi-periodic solutions to hierarchies of nonlinear integrable
equations and bilinear relations}

\author{A. Zabrodin\thanks{Skolkovo Institute of Science and
Technology, 143026, Moscow, Russia and
National Research University Higher School of
Economics,
20 Myasnitskaya Ulitsa, Moscow 101000, Russia, 
and NRC ``Kurchatov institute'', Moscow, Russia;
e-mail: zabrodin@itep.ru}}

\date{April 2023}
\maketitle

\vspace{-7cm} \centerline{ \hfill ITEP-TH-11/23}\vspace{7cm}

\begin{abstract}

This is a short review of the construction of 
quasi-periodic (alge\-bra\-ic-geo\-met\-ri\-cal)
solutions to hierarchies of nonlinear integrable equations. As is well known, the solutions
are expressed through Riemann's theta-functions associated with algebraic curves.
It is explained how solutions from this class can be treated within the framework of the
approach to the integrable hierarchies developed by the Kyoto school. 
Three representative examples are considered in detail: the Kadomtsev-Petviashvili
hierarchy, the 2D Toda lattice hierarchy and the B-version of the  Kadomtsev-Petviashvili
hierarchy. 

\end{abstract}

\end{titlepage}


%

\tableofcontents


\section{Introduction}

Exact quasi-periodic solutions to nonlinear integrable equations (soliton equations) 
are known 
to be expressed in terms of 
Riemann's theta-functions associated to algebraic curves (Riemann surfaces). 
For the first time the explicit
formulas were obtained for the $(1+1)$-dimensional Korteveg-de Vries equation,
see \cite{IM75,DMN76}. 

The unified approach to integration of more general $(2+1)$-dimensional Kadomtsev-Petviashvili
(KP) equation was developed by Krichever 
in 1977 \cite{Krichever77,Krichever77a}. It was shown that quasi-periodic
solutions are constructed starting from certain algebraic-geometrical data:
a smooth algebraic curve of finite genus 
with a marked point and a local parameter in the vicinity of this 
point. That is why such solutions are sometimes called ``algebraic-geometrical''. Explicit
formulas can be written in terms of the Riemann theta-functions, with the corresponding 
Riemann matrix being
the matrix of periods of holomorphic differentials (see also \cite{Dubrovin81,Dubrovin}).
Moreover, Novikov's conjecture (proved 
by Shiota in 1986 \cite{Shiota}) states that
exact theta-functional solutions to the KP 
equation solve the famous Schottky problem:
they give a characterization of Jacobian varieties among all abelian tori. 

The methods developed in \cite{Krichever77,Krichever77a} 
are applicable to other integrable
equations as well. Quasi-periodic solutions to 
the 2D Toda lattice equation were first
constructed in \cite{Krichever81}. Quasi-periodic solutions to the B-version 
of the KP hierarchy (BKP) were discussed in \cite{DJKM82a,N92}.

One of the basic objects in the construction of algebraic-geometrical solutions is the
Baker-Akhiezer function which can be regarded as a proper generalization of the exponential
function on the Riemann sphere to surfaces of nonzero genus. The key observation is that 
the Baker-Akhiezer function satisfies
an over-determined system of linear equations whose compatibility condition is equivalent
to the nonlinear equation. Remarkably, the Baker-Akhiezer function can be written explicitly
in terms of the Riemann theta functions and Abel map.  

A little bit later an algebraic 
analysis of nonlinear integrable equations was initiated in the works of the Kyoto 
school \cite{DJKM83,JimboMiwa}. The emphasis was made on the fact that the 
integrable equations form infinite hierarchies with infinitely many commuting flows. 
The corresponding analysis of the 2D Toda lattice hierarchy was given in \cite{UT84}.
A common solution to the whole hierarchy is provided by the tau-function which satisfies
a bilinear relation. The construction of algebraic-geometrical solutions can be 
extended to the hierarchy. It turns out that the corresponding tau-function is 
essentially the Riemann theta-function or Prym theta-function (for the BKP hierarchy). 

In this review we describe the construction of quasi-periodic solutions
to the KP, 2D Toda lattice and BKP hierarchies.
The main references are \cite{Krichever77,Krichever77a,Dubrovin81,Dubrovin,DJKM82a,handbook}.
In our presentation, the main attention is payed to the demonstration of how these 
theta-functional solutions can be embedded into the framework of the Kyoto school approach.
Our aim is to show that the equivalence of the two approaches is based on some nontrivial
facts about differentials on Riemann surfaces. 

Section 2 is a collection of necessary facts about 
Riemann surfaces and theta-functions
which we give without proofs. The material of section 2.2 is used only in section 5.
In sections 3, 4 and 5 we consider
quasi-periodic solutions to the KP, 2D Toda lattice and BKP hierarchies respectively.
We have tried to organize the presentation according to the 
following scheme: 
first we introduce the hierarchy by means of the bilinear 
relations for tau-function
following the approach of the Kyoto school, 
then give explicit expressions for the 
Baker-Akhiezer function and tau-function and, finally, show how they can be treated 
in the bilinear formalism.

\section{Algebraic-geometrical preliminaries}
\label{section:riemann}

Here we present the basic notions and facts about Riemann surfaces \cite{Springer,Fay},
theta-functions \cite{Mumford} and differentials on Riemann surfaces
which are necessary for the construction of quasi-periodic solutions to the 
integrable hierarchies.

\subsection{Arbitrary curves}

\label{section:arbitrary}

\paragraph{Matrix of periods and Jacobian.}
Let $\Gamma$ be a smooth compact algebraic curve (a Riemann surface) 
of genus $g$. We fix a canonical basis of
cycles ${\sf a}_{\alpha}, {\sf b}_{\alpha}$ ($\alpha =1, \ldots , g$) with the intersections
${\sf a}_{\alpha}\circ {\sf a}_{\beta}={\sf b}_{\alpha}\circ {\sf b}_{\beta}=0$,
${\sf a}_{\alpha}\circ {\sf b}_{\beta}=\delta_{\alpha \beta}$ and a basis of holomorphic 
differentials $d\omega_{\alpha}$ (abelian differentials of the first kind) 
normalized by the condition
$\displaystyle{\oint_{{\sf a}_{\alpha}}d\omega_{\beta}=\delta_{\alpha \beta}}$. 
The period matrix is defined as
\beq\label{qp1}
T_{\alpha \beta}=\oint_{{\sf b}_{\alpha}}d\omega_{\beta}, \qquad \alpha , \beta =1, \ldots , g.
\eeq
It is a symmetric matrix with positively defined imaginary part.

The Jacobian of the curve $\Gamma$ is the $g$-dimensional complex torus
\beq\label{qp4}
J(\Gamma )=\CC ^g /\{\vec N +T \vec M\},
\eeq
where $\vec N$, $\vec M$ are $g$-dimensional vectors with integer components.

\paragraph{The Abel map.}
Fix a point $Q_0\in \Gamma$ and define the Abel map $\vec A(P)$, $P\in \Gamma$
from $\Gamma$ to $J(\Gamma )$ as
\beq\label{qp3}
\vec A(P)=\vec \omega (P)=
\int_{Q_0}^P d \vec \omega , \qquad d\vec \omega =(d\omega_1, \ldots , d\omega_g ).
\eeq
The Abel map can be extended to the group of divisors ${\cal D}=n_1Q_1+\ldots +n_KQ_K$ as
\beq\label{qp3a}
\vec A({\cal D})=\sum_{i=1}^K n_i\int_{Q_0}^{Q_i} d \vec \omega =\sum_{i=1}^K
n_i\vec A(Q_i).
\eeq

\paragraph{The Riemann bilinear identity.}
Let ${\sf a}_{\alpha}$, ${\sf b}_{\alpha}$ be Jordan arcs 
which represent the canonical basis of cycles
intersecting transversally at a single point and let $$\tilde 
\Gamma =\Gamma \setminus \bigcup_{\alpha}
\Bigl ({\sf a}_{\alpha}\bigcup {\sf b}_{\alpha}\Bigr )$$ be a simply 
connected subdomain of $\Gamma$
(a fundamental domain). Integrals of any two meromorphic differentials $d\Omega '$,
$d\Omega ''$ satisfy the Riemann bilinear identity
\beq\label{r1}
\oint_{\p \tilde \Gamma}\Bigl (\int_{Q_0}^P d\Omega ' \Bigr )d\Omega ''(P)=
\sum_{\alpha =1}^g \left ( \oint_{{\sf a}_{\alpha}}d\Omega '\oint_{{\sf b}_{\alpha}}d\Omega ''-
\oint_{{\sf b}_{\alpha}}d\Omega '\oint_{{\sf a}_{\alpha}}d\Omega ''\right ).
\eeq
In particular, if $a$-periods of both differentials are equal to zero, we have
\beq\label{r2}
\oint_{\p \tilde \Gamma}\Bigl (\int_{Q_0}^P d\Omega ' \Bigr )d\Omega ''(P)=0.
\eeq

\paragraph{Abelian differentials of the second kind.}
Let $P_{\infty}\in \Gamma$ be a marked point and $k^{-1}$ a local parameter 
in a neighborhood of the marked point ($k=\infty$ at $P_{\infty}$). Let $d\Omega_j$ 
be differentials of the second kind with the only pole at $P_{\infty}$ of the form
$$
d\Omega_j = dk^j +O(k^{-2})dk, \quad k\to \infty
$$
normalized by the condition 
$\displaystyle{\oint_{{\sf a}_{\alpha}}d\Omega_{j}=0}$, and $\Omega_j$
be the (multi-valued) functions
$$
\Omega_j(P)=\int_{Q_0}^{P}d\Omega_j +q_j,
$$
where the constants $q_j$ are chosen in such a way that 
$\Omega_i(P)=k^i +O(k^{-1})$, namely,
\beq\label{qp3b}
\Omega_i(P)=k^i +\sum_{j\geq 1} \frac{1}{j}\, \Omega_{ij}k^{-j}.
\eeq
It follows from the Riemann bilnear identity that the matrix $\Omega_{ij}$ is symmetric:
$\Omega_{ij}=\Omega_{ji}$ (one should put $d\Omega '=d\Omega_i$, $d\Omega ''=d\Omega_j$
in (\ref{r2})).

Set
\beq\label{qp5}
U_j^{\alpha}=\frac{1}{2\pi i}\oint_{{\sf b}_{\alpha}}d\Omega_j, \qquad
\vec U_j =(U_j^{1}, \ldots , U_j^g).
\eeq
One can prove the following relation:
\beq\label{qp6a}
d\vec \omega =\sum_{j\geq 1}\vec U_j k^{-j-1}dk
\eeq
or
\beq\label{qp6}
\vec A(P)-\vec A(P_{\infty})=\int_{P_{\infty}}^P d\vec \omega =
-\sum_{j\geq 1}\frac{1}{j}\, \vec U_j k^{-j}
\eeq
(this follows from (\ref{r1}) if one puts $d\Omega '=d\omega_{\alpha}$,
$d\Omega ''=d\Omega_j$). 

Similarly, 
let $P_{0}\in \Gamma$ be another marked point with local parameter $k$ ($k(P_0)=0$).
Let $d\bar \Omega_j$ 
be differentials of the second kind with the only pole at $P_{0}$ of the form
$$
d\bar \Omega_j = dk^{-j} +O(1)dk, \quad k\to 0
$$
normalized by the condition $\displaystyle{\oint_{a_{\alpha}}d\bar \Omega_{j}=0}$, 
and 
$$
\bar \Omega_i(P)=\int_{Q_0}^{P}d\bar \Omega_i +\bar q_i=
k^{-i} +\sum_{j\geq 1} \frac{1}{j}\, \bar \Omega_{ij}k^{j}, 
\quad \bar \Omega_{ij}=\bar \Omega_{ji}.
$$
Set
\beq\label{qp5a}
V_j^{\alpha}=\frac{1}{2\pi i}\oint_{{\sf b}_{\alpha}}d\bar \Omega_j, 
\eeq
then the Riemann bilienar identity implies that
\beq\label{qp6b}
\vec A(P)-\vec A(P_{0})=
-\sum_{j\geq 1}\frac{1}{j}\, \vec V_j k^{j}.
\eeq

We will also need the expansions of the functions $\Omega_i(P)$, $\bar \Omega_i(P)$ near
the points $P_0$, $P_{\infty}$ respectively:
\beq\label{qp6c}
\begin{array}{l}
\displaystyle{
\Omega_i(P)=\Omega_i (P_0)+\sum_{j\geq 1}\frac{1}{j}\, \omega_{ij}k^j, \quad P\to P_0,}
\\ \\
\displaystyle{
\bar \Omega_i(P)=\bar \Omega_i (P_\infty )+
\sum_{j\geq 1}\frac{1}{j}\, \bar \omega_{ij}k^{-j}, \quad P\to P_\infty ,}
\end{array}
\eeq
and the Riemann bilienar identity implies that $\bar \omega_{ij}=\omega_{ji}$.

\paragraph{The Riemann theta-functions.}
The Riemann theta-function associated with the Riemann surface is defined by the 
absolutely convergent series
\beq\label{qp2}
\Theta(\vec z)=\Theta(\vec z|T)=
\sum_{\vec n \in \z ^{g}}e^{\pi i (\vec n, T\vec n)+2\pi i (\vec n, \vec z)},
\eeq
where $\vec z=(z_1, \ldots , z_g)$ and $\displaystyle{(\vec n, \vec z)=
\sum_{\alpha =1}^g n_{\alpha}z_{\alpha}}$.
It is an entire function with the following quasi-periodicity property:
\beq\label{qp2a}
\Theta (\vec z +\vec N +T\vec M )=\exp (-\pi i (\vec M, T\vec M)-
2\pi i (\vec M, \vec z)) \Theta (\vec z).
\eeq
More generally, one can introduce the theta-functions with characteristics
$\vec \delta'$, $\vec \delta'$:
\beq\label{char}
\Theta \left [\begin{array}{c}\vec \delta ' \\ \vec \delta^{''}
\end{array}\right ] (\vec z)
=\sum_{\vec n \in \z ^{g}}e^{\pi i (\vec n +\vec \delta ', 
T(\vec n +\vec \delta '))+2\pi i (\vec n+\vec \delta ', \vec z+\vec \delta '')}.
\eeq

\paragraph{Riemann's constants.}
Consider the function $f(P)=\Theta (\vec A(P)-\vec e)$ and assume that it is not 
identically zero. It can be shown that this function has $g$ zeros on $\Gamma$
at a divisor ${\cal D}=Q_1+ \ldots +Q_g$ and $\vec A( {\cal D})=\vec e -\vec K$,
where $\vec K=
(K_1, \ldots , K_g)$ is the 
vector of Riemann's constants
\beq\label{qp7}
K_{\alpha}= \pi i +\pi i T_{\alpha \alpha}-2\pi i
\sum_{\beta \neq \alpha}\oint_{a_{\beta}}\omega_{\alpha} (P)d\omega_{\beta}(P).
\eeq
In other words,
for any non-special effective divisor
${\cal D}=Q_1+\ldots +Q_g$ of degree $g$ the function
$$
f(P)=\Theta \Bigl (\vec A(P)-\vec A({\cal D}) -\vec K\Bigr )
$$
has exactly $g$ zeros at the points $Q_1, \ldots , Q_g$.  
Let ${\cal K}$ be the canonical class of divisors (the equivalence class of divisors
of poles and zeros of abelian differentials on $\Gamma$), then one can show that
\beq\label{qp8}
2\vec K=-\vec A({\cal K}).
\eeq
It is known that $\mbox{deg}\, {\cal K}=2g-2$. In particular, this means that 
holomorphic differentials have $2g-2$ zeros on $\Gamma$. 

\paragraph{The bi-differential.}
We also need the bi-differential $d_P d_Q \Omega (P, Q)$ such that it is symmetric in
$P, Q$, its
only singularity is a second order
pole at $P=Q$ and the integrals over $a$-cycles vanish. It is related to the
differentials $d\Omega _j$ as follows:
\beq\label{qp81}
\res_{Q=P_{\infty}} \Bigl (k^i(Q)d_P d_Q \Omega (P, Q)\Bigr ) =-d\Omega_i (P).
\eeq
The expansion in the local parameters is
\beq\label{qp82}
d_P d_Q \Omega (P, Q)=\Biggl ( \frac{1}{(k^{-1}(P)\! -\! k^{-1}(Q))^2}-
\sum_{i,j\geq 1} \Omega_{ij} k^{1-i}(P)k^{1-j}(Q)\Biggr ) dk^{-1}(P)dk^{-1}(Q).
\eeq
In fact this bi-differential can be expressed through the odd theta-function
$$\Theta_{*}(\vec z)=
\Theta \left [\begin{array}{c}\vec \delta '_{\rm o} \\ 
\vec \delta^{''}_{\rm o}
\end{array}\right ] (\vec z)
$$
($\Theta_{*}(\vec 0)=0$),
where $(\vec \delta '_{\rm o}, \vec \delta ''_{\rm o})$ 
is an odd half-integer theta-characteristics.
We have:
\beq\label{qp83}
d_P d_Q \Omega (P, Q)=d_P d_Q \log \Theta_{*}\Bigl (\vec A(P)-\vec A(Q)\Bigr ).
\eeq
Calculating the double integral
$\displaystyle{
\int_{P_{\infty}}^{P_1}\int_{Q_0}^{P_2}d_P d_Q \Omega (P, Q)}
$
in two ways (using first (\ref{qp82}) and then (\ref{qp83})), we obtain the equality
$$
\log \frac{(k^{-1}(P_1)-k^{-1}(P_2))
k^{-1}(Q_0)}{(k^{-1}(P_1)-k^{-1}(Q_0))k^{-1}(P_2)}
-\sum_{i,j\geq 1} \Omega_{ij} \frac{k^{-i}(P_1)k^{-j}(P_2)}{ij}+
\sum_{i,j\geq 1} \Omega_{ij} \frac{k^{-i}(P_1)k^{-j}(Q_0)}{ij}
$$
$$
=\log \frac{\Theta_{*}\Bigl (\vec A(P_2)-\vec A(P_1)\Bigr )
\Theta_{*}\Bigl (\vec A(P_\infty )\Bigr )}{\Theta_{*}\Bigl (\vec A(P_2)-
\vec A(P_\infty )\Bigr )\Theta_{*}\Bigl (\vec A(P_1 )\Bigr )}.
$$
Tending here $Q_0\to P_{\infty}$, we arrive at the relation
\beq\label{qp84}
\exp \Biggl (-\sum_{i,j\geq 1}\Omega_{ij}\frac{k_1^{-i}k_2^{-j}}{ij}\Biggr )=
\frac{C \Theta_{*}\Bigl ( \vec A(P_1)-\vec A(P_2)\Bigr )}{(k_1-k_2)
\Theta_{*}\Bigl ( \vec A(P_1)\! -\! \vec A(P_\infty )\Bigr )
\Theta_{*}\Bigl ( \vec A(P_2)\! -\! \vec A(P_\infty )\Bigr )},
\eeq
where $k_1=k(P_1), k_2=k(P_2)$ and
$$C=\sum_{\alpha =1}^{g}U_1^{\alpha} \Theta_{*, \alpha} (\vec 0), \qquad
\Theta_{*, \alpha}(\vec 0)=\frac{\p \Theta_{*}(\vec z)}{\p z_{\alpha}}\Biggr |_{\vec z=0}
$$
is a constant.


\paragraph{Abelian differential of the third kind.}
We will also need the abelian differentials of the third kind which are meromorphic differentials
with two simple poles with residues $\pm 1$ (dipole differentials). In particular, 
let $d\Omega_0$ be the differential having a simple pole 
at $P_0$ with residue $+1$ and 
another simple pole at $P_{\infty}$ with residue $-1$
with zero ${\sf a}$-periods. The expansions of the function 
$\Omega_0(P)$ near these points are
\beq\label{qp9}
\Omega_0(P)=\left \{
\begin{array}{l}
\displaystyle{
\log k +\Omega_{00}+
\sum_{j\geq 1}\frac{1}{j}\, \Omega_{0j}k^{-j}, \quad P\to P_{\infty},}
\\ \\
\displaystyle{
\log k +\bar \Omega_{00}+
\sum_{j\geq 1}\frac{1}{j}\, \bar \Omega_{0j}k^{j}, \quad P\to P_{0}.}
\end{array}
\right.
\eeq
Set
\beq\label{qp10}
U_0^{\alpha}=\frac{1}{2\pi i}\oint_{{\sf b}_{\alpha}}d\Omega_0.
\eeq
The following important relations are immediate consequences of the Riemann bilinear
identity:
\beq\label{qp11}
\vec A(P_0)-\vec A(P_{\infty})=U_0,
\eeq
\beq\label{qp12}
\Omega_{0i}=-\Omega_i(P_0), \qquad \bar \Omega_{0i}=\bar \Omega_i(P_{\infty}).
\eeq

Note that the function $\Omega_0(P)$ (\ref{qp9}) is given by
\beq\label{qp9a}
\Omega_0(P)=\log \frac{\Theta_{*}(\vec A(P)-
\vec A(P_0)}{\Theta_{*}(\vec A(P)-\vec A (P_{\infty}))}
\eeq
and the constant $\Omega_{00}$ in the expansion (\ref{qp9}) is the
following:
\beq\label{qp9b}
\Omega_{00}=-\log C +\log \Theta_{*}(\vec A(P_0)-\vec A(P_{\infty})).
\eeq

\paragraph{The holomorphic differential $d\zeta$.}
Tending
$k_1\to k_2$ in (\ref{qp84}), we get
\beq\label{qp85}
\exp \Biggl (-\sum_{i,j\geq 1}\Omega_{ij}\frac{k^{-i-j}}{ij}\Biggr )dk=
\frac{Cd\zeta}{\Theta_{*}^2\Bigl ( \vec A(P)\! -\! \vec A(P_\infty )\Bigr )},
\eeq
where $d\zeta$ is the holomorphic differential
\beq\label{qp86}
d\zeta =\sum_{\alpha =1}^g \Theta_{*, \alpha}(\vec 0)d\omega_{\alpha}.
\eeq
It is known (see \cite{Mumford}) that the differential $d\zeta$ has {\it double} zeros at some
$g-1$ points $R_1, \ldots , R_{g-1}$ while the function
$$
f_{*}(P)= \Theta_{*}\Bigl ( \vec A(P)\! -\! \vec A(P_\infty )\Bigr )
$$
has simple zeros at the same points $R_i$ and $P_{\infty}$. Therefore, the differential
in the right hand side of (\ref{qp85}) has the only (second order) pole at $P_{\infty}$
and no zeros. However, this differential is well-defined only on a covering of the curve
$\Gamma$ because it is not single-valued. 

The odd theta-function $\Theta_{*}$ and the differential $d\zeta$ depend on the choice
of the odd theta-characteristics but the prime form
\beq\label{prime}
E(P,Q)=\frac{\Theta_{*}(\vec A(P)-\vec A(Q))}{\sqrt{d\zeta (P)}\, \sqrt{d\zeta (Q)}}
\eeq
does not depend on it \cite{Fay,Mumford}. The bi-differential (\ref{qp83})
$d_Pd_Q\Omega(P,Q)=
d_Pd_Q \log E(P,Q)$ also does not depend on the choice of the odd theta-characteristics
(the $\frac{1}{2}$-differentials $\sqrt{d\zeta (P)}$, $\sqrt{d\zeta (Q)}$ cancel).

\paragraph{The Fay identity.}
At last, we mention the trisecant Fay identity \cite{Fay} satisfied by the function
$\Theta \Bigl (\vec z +\vec A(P)\Bigr )$:
\beq\label{Fay}
\begin{array}{c}
\Theta_{*}\Bigl (\vec A(P_1)\! -\! \vec A(P_2)\Bigr )
\Theta_{*}\Bigl (\vec A(P_3)\! -\! \vec A(P_4)\Bigr )
\Theta \Bigl (\vec z +\vec A(P_1)\! +\! \vec A(P_2)\Bigr )
\Theta \Bigl (\vec z +\vec A(P_3)\! +\! \vec A(P_4)\Bigr )
\\ \\
+\Theta_{*}\Bigl (\vec A(P_2)\! -\! \vec A(P_3)\Bigr )
\Theta_{*}\Bigl (\vec A(P_1)\! -\! \vec A(P_4)\Bigr )
\Theta \Bigl (\vec z +\vec A(P_2)\! +\! \vec A(P_3)\Bigr )
\Theta \Bigl (\vec z +\vec A(P_1)\! +\! \vec A(P_4)\Bigr )
\\ \\
+ \Theta_{*}\Bigl (\vec A(P_3)\!\! -\!\! \vec A(P_1)\Bigr )
\Theta_{*}\Bigl (\vec A(P_2)\!\! -\!\! \vec A(P_4)\Bigr )
\Theta \Bigl (\vec z \! +\! \vec A(P_1)\! +\! \vec A(P_3)\Bigr )
\Theta \Bigl (\vec z \! +\! \vec A(P_2)\! +\! \vec A(P_4)\Bigr )=0.
\end{array}
\eeq
Here $P_1, P_2, P_3, P_4$ are four arbitrary points on $\Gamma$ and $\vec z$ is an arbitrary
vector.

\subsection{Curves with holomorphic involution}

Assume now that the curve $\Gamma$ admits a 
holomorphic involution $\iota$ with two fixed
points $P_{\infty}$ and $P_0$. 
The Riemann-Hurwitz formula implies that genus $g$ is even:
$g=2g_0$. The curve $\Gamma$ is two-sheet covering of the factor-curve
$\Gamma_0=\Gamma /\iota$ of genus $g_0$. 
In the vicinity of $P_{\infty}$ we fix a local 
parameter $k^{-1}$ ($k^{-1}(P_{\infty})=0$) such that $k(\iota P)=-k(P)$. 

\paragraph{The Prym variety.}
We introduce a canonical basis of cycles ${\sf a}_{\alpha}, {\sf b}_{\alpha}$ 
on $\Gamma$  with the properties
$\iota {\sf a}_{\alpha}=-{\sf a}_{g_0+\alpha}$, $\iota {\sf b}_{\alpha}=-{\sf b}_{g_0+\alpha}$, 
$\alpha =1, \ldots , g_0$. Here and below
the sums like $\alpha +g_0$ for $\alpha =1, \ldots , g$ 
are understood modulo $g$, i.e., for example,
$(g_0+1)+g_0=1$.
Let $d\omega_{\alpha}$ be the corresponding 
basis of normalized holomorphic differentials, then $\iota^{*}d\omega_{\alpha}=
-d\omega_{g_0+\alpha}$, $\alpha =1, \ldots , g_0$. With this choice, the matrix
of periods enjoys the symmetry
\beq\label{inv0}
T_{\alpha \beta}=T_{\alpha +g_0, \beta +g_0}.
\eeq
Note that the symmetry property (\ref{inv0}) of 
the period matrix implies the relation
\beq\label{inv0a}
\Theta (\iota \vec u)=\Theta (\vec u) \quad \mbox{for any $u\in \CC^g$}.
\eeq

The Prym differentials are
defined as $d\upsilon_{\alpha}=d\omega_{\alpha}+d\omega_{g_0+\alpha}$
(here $\alpha =1, \ldots , g_0$); they are odd
with respect to the involution. The $g_0\times g_0$ matrix
\beq\label{inv1}
\Pi_{\alpha \beta}=\oint_{{\sf b}_{\alpha}}d\upsilon_{\beta}, 
\quad \alpha, \beta =1, \ldots , g_0
\eeq
is called the Prym matrix of periods. It is a symmetric matrix with positively 
defined imaginary part. 

The involution $\iota$ induces the involution of the Jacobian: $\iota u_{\alpha}
=-u_{\alpha +g_0}$, or, explicitly,
\beq\label{ind}
\iota (u_1, \ldots , u_g)=-(u_{g_0+1}, \ldots , u_g, u_1, \ldots , u_{g_0}).
\eeq
The Prym variety $Pr(\Gamma )\subset J(\Gamma )$ is a 
subvariety of the Jacobian
of $\Gamma$ defined by the condition $\iota (\vec u)=-\vec u$, i.e. it is the 
variety
$$
Pr(\Gamma )=
\Bigl \{ \vec u \in \CC^g \Bigr | \, \vec u =(u_1, 
\ldots , u_{g_0}, u_1, \ldots , u_{g_0})
\Bigr \}/(\ZZ^g + T\ZZ^g).
$$
It is isomorphic to the $g_0$-dimensional torus
$
Pr=\CC^{g_0}/((\ZZ^{g_0} + \Pi \ZZ^{g_0}).
$
This torus can be embedded into the Jacobian by the map
\beq\label{ind1}
\sigma (\vec u)=(u_1, \ldots , u_{g_0}, u_1, \ldots , 
u_{g_0})\in J(\Gamma ),
\quad \vec u =(u_1, \ldots , u_{g_0}),
\eeq
and the image is the Prym variety.

\paragraph{The Abel-Prym map.} 
The Abel-Prym map $\vec A^{\rm Pr}(P)$, $P\in \Gamma$
from $\Gamma$ to $Pr(\Gamma )$ is defined as
\beq\label{inv3}
\vec A^{\rm Pr}(P)=\vec \upsilon (P)=
\int_{P_0}^P d \vec \upsilon , \qquad d\vec \upsilon =
(d\upsilon_1, \ldots , d\upsilon_{g_0}),
\eeq
or
\beq\label{inv3a}
A^{{\rm pr}}_{\alpha}(P)=A_{\alpha}(P)+A_{g_0+\alpha}(P)=A_{\alpha}(P)-A_{\alpha}(\iota P).
\eeq
The Abel-Prym map can be extended to the group of divisors by linearity.
Note that for curves with involution the initial point of the Abel map is 
not arbitrary: it is $P_0$, the second fixed point of the involution. 
Note that according to our definitions 
$A_{\alpha}(\iota P)=-A_{g_0+\alpha}(P)$ that
agrees with the induced involution of the Jacobian (\ref{ind}).

\paragraph{Abelian differentials.}
The normalized abelian differentials of the second kind $d\Omega_j$ 
with poles at $P_{\infty}$ and corresponding 
abelian functions $\Omega_j$,
$$
\Omega_j(P)=\int_{P_0}^P d\Omega_j ,
$$
are defined in the same way as in section \ref{section:arbitrary}. 
Note that the initial point in the abelian integral defining the function
$\Omega_j$ is the second fixed point $P_0$.  The involution implies
\beq\label{inv4}
\iota^* d\Omega_j =(-1)^j d\Omega_j ,
\eeq
and for their $b$-periods we have
\beq\label{inv5}
U^{g_0+\alpha}_j = (-1)^{j+1} U^{\alpha}_j.
\eeq
Therefore, instead of (\ref{qp6}) it holds
\beq\label{inv6}
\vec A^{\rm Pr}(P)=-2\! \sum_{j\geq 1, \, {\rm odd}}
\frac{1}{j}\, \vec U_j k^{-j}.
\eeq

\paragraph{The Prym theta-function.}
The Prym theta-function is defined by the series
\beq\label{inv2}
\Theta_{\rm Pr}(\vec z)=\Theta_{\rm Pr}(\vec z|\Pi )=
\sum_{\vec n \in \z ^{g_0}}e^{\pi i (\vec n, \Pi \vec n)+2\pi i (\vec n, \vec z)}.
\eeq
Consider the function $f(P)=\Theta_{\rm Pr}(\vec A^{\rm Pr}(P)-\vec c)$
with $\vec c \in \CC^{g_0}$ and suppose that
it does not vanish identically. Then the zero divisor ${\cal D}$ of this function on 
$\Gamma$ is of degree $g$ and satisfies the relation
\beq\label{inv7}
\sigma (\vec c)=\vec A({\cal D})-\frac{1}{2} \vec A(P_{\infty })+\vec K.
\eeq

We also note the following important relation \cite{Fay}: for any $\vec u\in \CC^{g_0}$
it holds
\beq\label{inv8}
\begin{array}{c}
\Theta \Bigl (\sigma (\vec u)+\frac{1}{2}\vec A(P_{\infty})\Bigr )=
\Theta \Bigl (\sigma (\vec u)-\frac{1}{2}\vec A(P_{\infty})\Bigr )
=C \Bigl (\Theta_{\rm Pr}(\vec u)\Bigr )^2,
\end{array}
\eeq
i.e., the Riemann theta-function on this part of the Jacobian is a full square
and the square root is, up to a constant multiplier, the Prym theta-function.

\paragraph{Analogue of (\ref{qp85}).} Let us consider the bi-differential
\beq\label{an0}
d_Pd_Q \, \omega (P,Q)=d_Pd_Q \Omega (P,Q)-d_Pd_Q \Omega (P,\iota Q)=
d_Pd_Q \log \frac{E(P,Q)}{E(P, \iota Q)}.
\eeq
Note that for curves with involution the double sum in (\ref{qp82}) contains only terms
in which $i,j$ have the same parity. Taking this into account and recalling that
$k(\iota Q)=-k(Q)$, we can write
\beq\label{an1}
\begin{array}{l}
\displaystyle{
d_Pd_Q \, \omega (P,Q)=\left ( \frac{1}{(k^{-1}(P)-k^{-1}(Q))^2}+
\frac{1}{(k^{-1}(P)+k^{-1}(Q))^2}\right. }
\\ \\
\displaystyle{\vphantom{\frac{1}{(k^{-1}(P)-k^{-1}(Q))^2}}\phantom{aaaaaaa}
\left.
-2\sum_{i,j \,\,{\rm odd}}\Omega_{ij}k^{1-i}(P)k^{1-j}(Q)
\right )dk^{-1}(P)dk^{-1}(Q)}.
\end{array}
\eeq
Calculating the double integral $\displaystyle{\int_{Q_0'}^{P}\int_{Q_0}^{Q}d_Pd_Q \omega }$
in two ways (using (\ref{an0}) and (\ref{an1})) and equating the results, we obtain, in the
limit $Q_0\to P_{\infty}$, $Q_0'\to P_{\infty}$:
\beq\label{an2}
\log \frac{k^{-1}(P)-k^{-1}(Q)}{k^{-1}(P)+k^{-1}(Q)}-2\sum_{i,j \,\, {\rm odd}}
\Omega_{ij}\, \frac{k^{-i}(P)k^{-j}(Q)}{ij}=
\log \frac{E(P,Q)}{E(P,\iota Q)}.
\eeq
Exponentiating and tending $Q\to P$, we obtain, using the expression (\ref{prime}) for the
prime form:
\beq\label{an3}
\exp \Bigl (-2\sum_{i,j \,\, {\rm odd}}
\Omega_{ij}\, \frac{k^{-i-j}}{ij}\Bigr )\frac{dk}{k}=-
\frac{2\sqrt{d\zeta (P)d\zeta (\iota P)}}{\Theta_{*}(\vec A(P)-\vec A(\iota P))}.
\eeq
This formula is an analogue of (\ref{qp85}).
The denominator has simple zeros at the points $P_0, P_{\infty}$ and at some points
$R_i, \iota R_i$, $i=1, \ldots , g-1$. The latter $2g-2$ poles are canceled by zeros of the 
holomorphic differential in the numerator.

\section{The KP hierarchy}

\subsection{Bilinear relations for the KP hierarchy}

The set of independent variables in the KP hierarchy is the infinite set of
(in general complex) ``times'' ${\bf t}=\{t_1, t_2, t_3, \ldots \, \}$. 
The universal dependent variable is the tau-function $\tau =\tau ({\bf t})$.
All equations of the hierarchy are encoded in the bilinear relation
\beq\label{kp1}
\oint_{C_{\infty}}e^{\xi ({\bf t}-{\bf t}', z)}
\tau ({\bf t}-[z^{-1}])\tau ({\bf t}'+[z^{-1}])\, dz =0
\eeq
valid for arbitrary ${\bf t}$, ${\bf t}'$, where
$$
{\bf t}\pm [z^{-1}]=\Bigl \{ t_1 \pm z^{-1}, t_2\pm \frac{1}{2}z^{-2},
t_3\pm \frac{1}{3}z^{-3}, \ldots \Bigr \}
$$
and 
\beq\label{kp2}
\xi ({\bf t}, z)=\sum_{j\geq 1} t_j z^j.
\eeq
The contour $C_{\infty}$ in (\ref{kp1}) is a big circle encircling $\infty$. 
Note that the simplest (trivial) solution of (\ref{kp1}) is $\tau ({\bf t})=1$.

Let us set ${\bf t}-{\bf t}'=[z_1^{-1}]+[z_2^{-1}]+[z_3^{-1}]-[z_4^{-1}]$ in (\ref{kp1}). 
The residue calculus yields 
\beq\label{kp3}
\begin{array}{l}
(z_1-z_2)(z_3-z_4)\tau ({\bf t}-[z_1^{-1}]-[z_2^{-1}])
\tau ({\bf t}-[z_3^{-1}]-[z_4^{-1}])
\\ \\
\phantom{aaaaa}
+(z_2-z_3)(z_1-z_4)\tau ({\bf t}-[z_2^{-1}]-[z_3^{-1}])
\tau ({\bf t}-[z_1^{-1}]-[z_4^{-1}])
\\ \\
\phantom{aaaaaaaaaa}
+ (z_3-z_1)(z_2-z_4)\tau ({\bf t}-[z_1^{-1}]-[z_3^{-1}])
\tau ({\bf t}-[z_2^{-1}]-[z_4^{-1}])=0
\end{array}
\eeq
which is the Hirota-Miwa equation for the tau-function.
It is a generating equation for the differential equations of the hierarchy.
They are obtained by expanding it in negative powers of
$z_1, z_2, z_3, z_4$.
In the limit $z_4\to \infty$, $z_3\to \infty$ equation (\ref{kp3}) becomes
\beq\label{kp4}
\begin{array}{l}
\displaystyle{
\p_{t_1}\log \frac{\tau\Bigl ({\bf t}+
[z_1^{-1}]-[z_2^{-1}]\Bigr )}{\tau ({\bf t})}
=
(z_2-z_1)\left (\frac{\tau \Bigl ({\bf t}+
[z_1^{-1}]\Bigr )\tau \Bigl ({\bf t}-
[z_2^{-1}]\Bigr )}{\tau ({\bf t})
\tau \Bigl ({\bf t}+
[z_1^{-1}]-[z_2^{-1}]\Bigr )}-1\right )}.
\end{array}
\eeq

Important objects of the theory are the wave function $\psi ({\bf t}, z)$ and its dual
$\psi^{\dag}({\bf t}, z)$. They are expressed through the tau-function as follows:
\beq\label{kp5}
\psi ({\bf t}, z)=e^{\xi ({\bf t}, z)}\frac{\tau ({\bf t}-[z^{-1}])}{\tau ({\bf t})},
\eeq
\beq\label{kp6}
\psi^{\dag} ({\bf t}, z)=e^{-\xi ({\bf t}, z)}\frac{\tau ({\bf t}+[z^{-1}])}{\tau ({\bf t})}.
\eeq
Here it is implied that the spectral parameter 
$z$ belongs to some neighborhood of $\infty$. In terms of the wave functions, 
the bilinear relation (\ref{kp1}) acquires the form
\beq\label{kp7}
\oint_{C_{\infty}}\psi ({\bf t}, z)\psi^{\dag}({\bf t}', z)\, dz =0.
\eeq

\subsection{The one-point Baker-Akhiezer functions}

\label{section:baker}

The (one-point) Baker-Akhiezer function 
$\Psi =\Psi ({\bf t}, P)$ is a function
on the curve $\Gamma$ of genus $g$ ($P$ is a point of $\Gamma$)
which is analytic everywhere except at the marked point $P_{\infty}$, where
it has an essential singularity of the prescribed form and $g$ simple poles at some 
points $Q_1, \ldots , Q_g \in \Gamma$. We denote the divisor of poles of $\Psi$ by 
$${\cal D}=Q_1+ \ldots + Q_g$$ 
and assume that it is non-special. 
The standard argument shows that the linear space of Baker-Akhiezer functions is
one-dimensional. In other words, the Baker-Akhiezer function is unique up to a common
multiplier.
In the vicinity of $P_{\infty}$
the Baker-Akhiezer function as a function of the local parameter
$k^{-1}$ has the form 
\beq\label{ba1}
\Psi ({\bf t}, P)=e^{\xi ({\bf t}, k)}
\Bigl (1+\xi_1 ({\bf t})k^{-1} +\xi_2({\bf t}) k^{-2} +\ldots \Bigr ), \quad P\to P_{\infty}
\eeq
(we assume that only a finite
number of the times $t_j$ are different from zero). The above mentioned common
multiplier is fixed by the condition $\displaystyle{\lim_{P\to P_\infty}\Bigl (
e^{-\xi ({\bf t}, k)}\Psi ({\bf t}, P)\Bigr )=1}$.

Let $d\Omega$ be the abelian differential of the second kind 
with the only (second order) pole at $P_{\infty}$
and $g$ simple zeros at the points of the divisor ${\cal D}$. Such a differential is unique
up to a common constant multiplier. It has other $g$ zeros at some points
$Q_1^{\dag}, \ldots , Q_g^{\dag}$. Let 
$${\cal D}^{\dag}=Q_1^{\dag}+ \ldots + Q_g^{\dag}$$
be the divisor of these zeros, then we have the relation
\beq\label{ba2}
{\cal D}+{\cal D}^{\dag}={\cal K}+2P_{\infty}
\eeq
which means that
\beq\label{ba3}
\vec A({\cal D})+\vec A({\cal D}^{\dag})+2\vec K -2\vec A(P_{\infty})=0
\eeq
in the Jacobian. One can introduce the
adjoint (dual) Baker-Akhiezer function 
$\Psi^{\dag}$ by the requirement that it
has the divisor of poles ${\cal D}^{\dag}$ and 
in the vicinity of $P_{\infty}$ it behaves as
\beq\label{ba4}
\Psi^{\dag}({\bf t}, P) = e^{-\xi ({\bf t}, k)}
\Bigl (1+\xi_1^{\dag}({\bf t}) k^{-1} +\xi_2^{\dag}({\bf t}) k^{-2} +\ldots \Bigr ),
\quad P\to P_{\infty}.
\eeq

The differential $\Psi ({\bf t}, P)\Psi^{\dag}({\bf t}', P)d\Omega (P)$
is holomorphic 
everywhere on $\Gamma$ except the point $P_{\infty}$ (poles of the Baker-Akhiezer
functions are canceled by zeros of $d\Omega$). Therefore, the residue at this point
is equal to zero and we have
\beq\label{ba7}
\oint_{C_{\infty}} \! \Psi ({\bf t}, P)\Psi^{\dag}({\bf t}', P)d\Omega (P) =0
\eeq
for all ${\bf t}, {\bf t}'$, where $C_{\infty}$ is a small contour around the point $P_{\infty}$.

Using the Riemann's theta-function and the Abel map, one can write the Baker-Akhiezer
function in the explicit form:
\beq\label{ba5}
\begin{array}{l}
\displaystyle{
\Psi ({\bf t},P)= \exp \Bigl (\sum_{j\geq 1}t_j \Omega_j (P)\Bigr )
\frac{\Theta \Bigl (\vec A(P)\! +\!\sum\limits_{j\geq 1}\vec U_j t_j \!-\!
\vec A({\cal D}) \!-\! \vec K\Bigr )\Theta \Bigl (\vec A({\cal D}) \! +\!
\vec K\! -\! \vec A(P_{\infty})\Bigr )}{\Theta \Bigl (\vec A(P)\! -\! 
\vec A({\cal D}) \! -\! \vec K\Bigr )\Theta \Bigl (\sum\limits_{j\geq 1}\vec U_j t_j
\! -\! \vec A({\cal D}) \! -\!
\vec K\! +\! \vec A(P_{\infty})\Bigr ) }},
\end{array}
\eeq
where we use the notation introduced in section \ref{section:riemann}. In (\ref{ba5}) it
is assumed that
the path of integration in the Abel map and $\Omega_j$ is the same. 
This function is normalized in such a way that $\Psi (0, P)=1$. 
The corresponding expression for the dual Baker-Akhiezer function is
\beq\label{ba5a}
\begin{array}{l}
\displaystyle{
\Psi^{\dag} ({\bf t}, 
P)= \exp \Bigl (-\! \sum_{j\geq 1}t_j \Omega_j (P)\Bigr ) 
\frac{\Theta \Bigl (\vec A(P) \!-\! \sum\limits_{j\geq 1}\vec U_j t_j  \!-\!
\vec A({\cal D}^{\dag}) \! -\! \vec K\Bigr )\Theta \Bigl (\vec A({\cal D}^{\dag}) \! +\! 
\vec K \!-\! \vec A(P_{\infty})\Bigr )}{\Theta \Bigl (\vec A(P)\! -\! 
\vec A({\cal D}^{\dag}) \! -\! \vec K\Bigr )\Theta \Bigl (\sum\limits_{j\geq 1}\vec U_j t_j
\! +\! \vec A({\cal D}^{\dag}) \! +\! 
\vec K\! -\! \vec A(P_{\infty})\Bigr ) }}.
\end{array}
\eeq
Using the relation (\ref{ba3}), one can rewrite it in the form
\beq\label{ba6}
\begin{array}{l}
\displaystyle{
\Psi^{\dag} ({\bf t},P)= \exp \Bigl (-\! \sum_{j\geq 1}t_j \Omega_j (P)\Bigr )}
\\ \\
\phantom{aaaaa}\displaystyle{\times
\frac{\Theta \Bigl (\vec A(P)\! -\! \sum\limits_{j\geq 1}\vec U_j t_j \! +\! 
\vec A({\cal D}) \! +\! \vec K \! -\! 2\vec A(P_{\infty})\Bigr )\Theta \Bigl (\vec A({\cal D}) 
\! +\! 
\vec K\! -\! \vec A(P_{\infty})\Bigr )}{\Theta \Bigl (\vec A(P)\! +\! 
\vec A({\cal D}) \! +\! \vec K-2\vec A(P_{\infty})\Bigr )
\Theta \Bigl (\sum\limits_{j\geq 1}\vec U_j t_j
\! -\! \vec A({\cal D}) \! -\! 
\vec K\! +\! \vec A(P_{\infty})\Bigr ) }}.
\end{array}
\eeq

The standard argument shows that the Baker-Akhiezer function satisfies the linear 
equation
\beq\label{lin1}
\p_{t_2}\Psi =\p_x^2 \Psi +u\Psi , \quad x=t_1
\eeq
(which has the form of the non-stationary Schr\"odinger equation), where
\beq\label{lin2}
u(x)=-\p_x\xi_1.
\eeq

\subsection{The tau-function}

The tau-function is defined as
\beq\label{tau1}
\tau ({\bf t})=\exp \Bigl ( -
\frac{1}{2}\sum_{i,j\geq 1}\Omega_{ij}t_it_j\Bigr )\,
\Theta \Bigl (\sum_{j\geq 1}\vec U_j t_j +\vec Z \Bigr ), 
\eeq
where the constant vector $\vec Z$ is parametrized through the divisor ${\cal D}$ as
\beq\label{tau1a}
\vec Z=-\vec A({\cal D})-\vec K
+\vec A(P_{\infty}).
\eeq
Equation (\ref{qp6}) implies that 
$$
\begin{array}{c}
\displaystyle{
\Theta \Bigl (\sum_{j\geq 1}\vec U_j (t_j \mp \frac{1}{j}\, k^{-j}) +\vec Z \Bigr )
=\Theta \Bigl (\pm \vec A(P)+\sum_{j\geq 1}\vec U_j t_j  +\vec Z \mp \vec A(P_{\infty}) \Bigr ),}
\end{array}
$$
so we see that the Baker-Akhiezer functions (\ref{ba5}), (\ref{ba6})
are connected with the tau-function in the same way as the wave functions (see 
(\ref{kp5}), (\ref{kp6})). 
Namely, in a neighborhood of $P_{\infty}$ we have
\beq\label{tau2}
\Psi({\bf t}, P) = C(k)\psi ({\bf t}, k), \quad \Psi^{\dag}
({\bf t}, P) = C^{\dag}(k)\psi^{\dag} ({\bf t}, k),
\eeq
where 
\beq\label{tau2a}
C(k)=\frac{\tau (0)}{\tau (-[k^{-1}])}, \quad
C^{\dag}(k)=\frac{\tau (0)}{\tau ([k^{-1}])}.
\eeq

A simple calculation shows that
\beq\label{tau6a}
\begin{array}{l}
\displaystyle{
\frac{\tau ({\bf t}-[k_1^{-1}]-[k_2^{-1}])\tau ({\bf t})}{\tau ({\bf t}-[k_1^{-1}])
\tau ({\bf t}-[k_2^{-1}])}=
\exp \Biggl (-\! \sum_{i,j\geq 1}\Omega_{ij}\frac{k_1^{-i}k_2^{-j}}{ij}\Biggr )}
\\ \\
\phantom{aaaaaaaaaaaaaa}\displaystyle{\times \,
\frac{\Theta \Bigl (\vec A(P_1)+\vec A(P_2) +\sum\limits_{j\geq 1}\vec U_j t_j +
\vec Z \Bigr )\Theta \Bigl (\sum\limits_{j\geq 1}\vec U_j t_j +
\vec Z \Bigr )}{\Theta \Bigl (\vec A(P_1)+\sum\limits_{j\geq 1}\vec U_j t_j +
\vec Z \Bigr )\Theta \Bigl (\vec A(P_2) +\sum\limits_{j\geq 1}\vec U_j t_j +
\vec Z \Bigr )}}.
\end{array}
\eeq
Using (\ref{qp84}), it is straightforward to check that the tau-function 
(\ref{tau1}) satisfies the Hirota-Miwa equation (\ref{kp3}).
It appears to be equivalent to the
Fay identity (\ref{Fay}).

It is instructive to compare equation (\ref{ba7}) and the standard bilinear
relation for the tau-function (\ref{kp1}) or (\ref{kp7}) (here we follow \cite{KZ21}).
Comparing the two equations (\ref{ba7}) and (\ref{kp1}), we conclude that 
they coincide if it holds
\beq\label{tau5}
\begin{array}{c}
\displaystyle{
d\Omega =\frac{\tau (-[k^{-1}])\tau ([k^{-1}])}{\tau^2(0)}}\, dk
\\ \\
\displaystyle{
=
\frac{\Theta \Bigl (\vec A (P) -\vec A({\cal D}) -\vec K\Bigr )
\Theta \Bigl (\vec A (P) -\vec A({\cal D}^{\dag}) -
\vec K\Bigr )}{\Theta^2 \Bigl (\vec A({\cal D}) +
\vec K-\vec A(P_{\infty})\Bigr )}
\exp \Bigl (-\! \sum_{i,j\geq 1}\Omega_{ij}\, \frac{k^{-i-j}}{ij}\Bigr ) \, dk.}
\end{array}
\eeq
Using (\ref{qp85}), we can rewrite this as
\beq\label{tau7}
d\Omega =
C\frac{\Theta \Bigl (\vec A (P) -\vec A({\cal D}) -\vec K\Bigr )
\Theta \Bigl (\vec A (P) -\vec A({\cal D}^{\dag}) -
\vec K\Bigr )}{\Theta_{*}^2\Bigl ( \vec A(P)\! -\! \vec A(P_\infty )\Bigr )} \, d\zeta ,
\eeq
where the holomorphic differential $d\zeta$ is given by (\ref{qp86}). Its properties
(see \cite{Mumford} and section \ref{section:riemann}) imply that the differential 
in the right hand side is a well-defined meromorphic differential on $\Gamma$ with the
only second order pole at $P_{\infty}$ and $2g$ zeros at the points of the divisors
${\cal D}$, ${\cal D}^{\dag}$. Therefore, it has all the properties of the 
differential $d\Omega$ and hence must be proportional to it. 

\subsection{On the choice of the local parameter}

The tau-function (\ref{tau1}) provides a quasi-periodic solution to the KP hierarchy
in the Hirota form. The solution depends on the algebraic curve $\Gamma$ and on the choice
of the local parameter $k^{-1}$ in a neighborhood of the marked point $P_{\infty}$. The latter
dependence can be easily controlled. 

Let $\tilde k^{-1}$ be another local parameter:
$$
k=\tilde k +a_0 +a_1 \tilde k^{-1}+a_2 \tilde k^{-2}+\ldots , \qquad k\to \infty .
$$
Then we can write
\beq\label{lp0a}
dk^i = \sum_{l=1}^i a_{il} d\tilde k^l +d(k^i)_{-}
\eeq
with some triangular 
matrix ${\sf a}=(a_{ij})$ given by
\beq\label{lp0}
a_{il}=\res_{\infty} \Bigl (k^i \tilde k^{-l}d\tilde k/\tilde k\Bigr ),
\eeq 
where the residue is defined as $\res_{\infty}(k^{-j} dk)=\delta_{1j}$.
In (\ref{lp0a}), 
$(\ldots )_{\pm}$ means projection 
on the positive (negative) power part of the series with respect to $\tilde k$. 
This implies the following relation between the differentials $d\Omega _i$ and the
differentials $d\tilde \Omega_i$ defined in a similar way and with the same normalization 
with the help of the local parameter $\tilde k^{-1}$:
\beq\label{lp1}
d\Omega_i=\sum_{l=1}^i a_{il} d\tilde \Omega_l.
\eeq
Then a similar relation for their periods holds:
\beq\label{lp2}
U_j^{\alpha}=\sum_{l=1}^i a_{il} \tilde U_l^{\alpha} \quad \mbox{or} \quad
{\bf U}^{\alpha}={\sf a}\tilde {\bf U}^{\alpha},
\eeq
where ${\bf U}^{\alpha}$ is the half-infinite column 
vector with components $U_j^{\alpha}$.

Let us define the times $\tilde t_j$ by the relation
\beq\label{lp3}
\sum_{j\geq 1} t_j dk^j=\sum_{j\geq 1} \tilde t_j d\tilde k^j+O(\tilde k^{-2})d\tilde k.
\eeq
This relation means that 
the times $\tilde t_j$ are linear combinations of the $t_j$'s:
\beq\label{lp4}
\tilde t_j=\sum_{l\geq j}t_l a_{lj} \quad \mbox{or} \quad
\tilde {\bf t}={\bf t}{\sf a}.
\eeq
Using (\ref{lp2}), we see that $\displaystyle{\sum_{j\geq 1}t_j U_{j}^{\alpha}}$ does not change
when we change the local parameter:
\beq\label{lp5}
\sum_{j\geq 1}t_j U_{j}^{\alpha}={\bf t}{\bf U}^{\alpha}=
\tilde {\bf t}{\sf a}^{-1}{\sf a}\tilde {\bf U}^{\alpha}=
\sum_{j\geq 1}\tilde t_j \tilde U_{j}^{\alpha},
\eeq
so the theta-function in (\ref{tau1}) remains the same. 

Next, we can write:
\beq\label{lp6}
\tilde {\bf t}\tilde \Omega \tilde {\bf t}={\bf t}\Omega {\bf t}+{\bf t}\Delta {\bf t},
\qquad
\Delta = {\sf a} \tilde \Omega {\sf a}^T -\Omega ,
\eeq
where $\Omega$ is the matrix $\Omega_{ij}$ (see (\ref{qp3b})) and $\Delta$ is some matrix
which we are going to find explicitly. For this purpose, we write
$$
d\tilde \Omega_i=d\tilde k^i -\sum_{l\geq 1}\tilde \Omega_{il}\tilde k^{-l-1}d\tilde k
$$
and act to the both sides of this equation by the matrix ${\sf a}$. We get:
$$
\sum_{l=1}^i a_{il}d\tilde \Omega_{l}=d\Omega_i =dk^i -d(k^i)_{-}-
\sum_{j\geq 1}\sum_{l=1}^i a_{il}\tilde \Omega_{lj}\tilde k^{-j-1}d\tilde k.
$$
On the other hand,
$$
d \Omega_i=d k^i -\sum_{l\geq 1} \Omega_{il} k^{-l-1}d k,
$$
so we have the equality
$$
d(k^i)_{-}+
\sum_{j\geq 1}\sum_{l=1}^i a_{il}\tilde \Omega_{lj}\tilde k^{-j-1}d\tilde k =
\sum_{l\geq 1} \Omega_{il} k^{-l-1}d k.
$$
Multiplying it by $k^m$ and taking the residue, we obtain:
\beq\label{lp7}
\sum_{l,j}a_{il}\tilde \Omega_{lj}a_{mj}-\Omega_{im}=-\res_{\infty}\Bigl (
k^m d(k^i)_{-}\Bigr )=\Delta_{im}.
\eeq
It is easy to see that the matrix $\Delta_{ij}$ is symmetric and can be equivalently 
written as
\beq\label{lp8}
\Delta_{ij}=-\res_{\infty}\Bigl ((k^i)_{+}dk^j\Bigr ).
\eeq
Therefore, we arrive at the relation between the tau-functions (\ref{tau1}) and
$\tilde \tau (\tilde {\bf t})$ given by the same formula, where all quantities 
(except $\vec Z$) go with tilde:
\beq\label{lp9}
\tilde \tau (\tilde {\bf t})=\exp \Bigl (-\frac{1}{2}
\sum_{i,j\geq 1}\Delta_{ij}t_it_j\Bigr )\, \tau ({\bf t}).
\eeq
We see that the theta-function multiplier remains invariant and only the 
quadratic form changes.

\section{2D Toda lattice hierarchy}

\subsection{Bilinear relations for the Toda lattice hierarchy}

\label{section:todabilinear}

The independent variables in the 2D Toda lattice hierarchy are two infinite sets 
of complex variables
${\bf t}=\{t_1, t_2, t_3, \ldots \, \}$ (``positive times''),
$\bar {\bf t}=\{\bar t_1, \bar t_2, \bar t_3, \ldots \, \}$ (``negative times'')
and an integer variable $n\in \ZZ$ (``zeroth time''). The dependent variable
is the tau-function $\tau_n({\bf t}, \bar {\bf t})$. All equations of the hierarchy
are encoded in the bilinear relation
\beq\label{toda1}
\begin{array}{l}
\displaystyle{\oint_{C_{\infty}}\! z^{n-n'-1}\, 
e^{\xi ({\bf t}-{\bf t}', z)}\tau_n \Bigl ({\bf t}-[z^{-1}], \bar {\bf t}\Bigr )
\tau_{n'+1} \Bigl ({\bf t}'+[z^{-1}], \bar {\bf t}'\Bigr )\, dz}
\\ \\
\phantom{aaaaaaaaa}
\displaystyle{=\oint_{C_0}\! z^{n-n'-1}\, 
e^{\xi (\bar {\bf t}-\bar {\bf t}', z^{-1})}\tau_{n+1}\Bigl ({\bf t}, \bar {\bf t}-[z]\Bigr )
\tau_{n'}\Bigl ({\bf t}', \bar {\bf t}'+[z]\Bigr )\, dz}
\end{array}
\eeq
valid for all $n, {\bf t}, \bar {\bf t}$ and $n', {\bf t}', \bar {\bf t}'$, 
where the contour $C_{\infty}$ is the same as 
in (\ref{kp1}) and $C_{0}$ is a small circle around $0$. Note that
\beq\label{simplest}
\tau_n ({\bf t}, \bar {\bf t})=\exp \Bigl (\sum_{j\geq 1}jt_j\bar t_j\Bigr )
\eeq
is the simplest (trivial) solution. 

Clearly, the KP hierarchy can be embedded into the 
2D Toda lattice hierarchy by fixing $n=0$, $\bar {\bf t}=0$. 
Fixing $\bar {\bf t}=\bar {\bf t}'=0$,
one obtains what is called the modified KP (mKP) hierarchy with the bilinear relation
\beq\label{mkp}
\oint_{C_{\infty}}\! z^{m-1}\, 
e^{\xi ({\bf t}-{\bf t}', z)}\tau_n \Bigl ({\bf t}-[z^{-1}]\Bigr )
\tau_{n-m+1} \Bigl ({\bf t}'+[z^{-1}]\Bigr )\, dz=
\delta_{m,0}\tau_{n+1}({\bf t})\tau_{n}({\bf t}'), \quad m\geq 0.
\eeq

Let us point out some important consequences of the bilinear relation.
If we set $n-n'=1$, ${\bf t}-{\bf t}'=[a^{-1}]$, $\bar {\bf t}-\bar {\bf t}'=[b^{-1}]$, then
the residue calculus yields
\beq\label{toda2}
\begin{array}{c}
\tau_n\Bigl ({\bf t}-[a^{-1}], \bar {\bf t}\Bigr )
\tau_n\Bigl ({\bf t}, \bar {\bf t}-[b^{-1}]\Bigr )-
\tau_n({\bf t}, \bar {\bf t})
\tau_n\Bigl ({\bf t}-[a^{-1}], \bar {\bf t}-[b^{-1}]\Bigr )
\\ \\
=(ab)^{-1} \tau_{n+1}\Bigl ({\bf t}, \bar {\bf t}-[b^{-1}]\Bigr )
\tau_{n-1}\Bigl ({\bf t}-[a^{-1}], \bar {\bf t}\Bigr ).
\end{array}
\eeq
Differentiating (\ref{toda1}) with respect to $t_1$ and then setting
$n=n'$, ${\bf t}-{\bf t}'=[a^{-1}]$, $\bar {\bf t}-\bar {\bf t}'=[b^{-1}]$, we get:
\beq\label{toda3}
\p_{t_1}\log \frac{\tau_{n+1}\Bigl ({\bf t}, \bar {\bf t}-[b^{-1}]\Bigr )}{\tau_n
\Bigl ({\bf t}-[a^{-1}], \bar {\bf t}\Bigr )}=
a\left (1-\frac{\tau_n ({\bf t}, \bar {\bf t})\tau_{n+1}
\Bigl ({\bf t}-[a^{-1}], \bar {\bf t}-[b^{-1}]\Bigr )}{\tau_n\Bigl ({\bf t}-[a^{-1}], 
\bar {\bf t}\Bigr )\tau_{n+1}\Bigl ({\bf t}, \bar {\bf t}-[b^{-1}]\Bigr )}\right ).
\eeq

Important objects of the theory are the wave functions $\psi_n$, $\bar \psi_n$ 
and their dual $\psi^{\dag}_n$, $\bar \psi^{\dag}_n$. They are expressed through the 
tau-function as follows:
\beq\label{toda4}
\psi_n ({\bf t}, \bar {\bf t}, z)=z^ne^{\xi ({\bf t}, z)}
\frac{\tau_n\Bigl ({\bf t}-[z^{-1}], \bar {\bf t}\Bigr )}{\tau_n ({\bf t}, \bar {\bf t})},
\eeq
\beq\label{toda4a}
\bar \psi_n ({\bf t}, \bar {\bf t}, z)=z^ne^{\xi (\bar {\bf t}, z^{-1})}
\frac{\tau_{n+1}\Bigl ({\bf t}, \bar {\bf t}-[z]\Bigr )}{\tau_n ({\bf t}, \bar {\bf t})},
\eeq
\beq\label{toda4b}
\psi_n^{\dag} ({\bf t}, \bar {\bf t}, z)=z^{-n}e^{-\xi ({\bf t}, z)}
\frac{\tau_{n+1}\Bigl ({\bf t}+[z^{-1}], \bar {\bf t}\Bigr )}{\tau_{n+1} ({\bf t}, \bar {\bf t})},
\eeq
\beq\label{toda4c}
\bar \psi_n^{\dag} ({\bf t}, \bar {\bf t}, z)=z^{-n}e^{-\xi (\bar {\bf t}, z^{-1})}
\frac{\tau_{n}\Bigl ({\bf t}, \bar {\bf t}+[z]\Bigr )}{\tau_{n+1} ({\bf t}, \bar {\bf t})}.
\eeq
In terms of the wave functions, the bilinear relation reads
\beq\label{toda5}
\oint_{C_{\infty}}\! \frac{dz}{z}\, \psi_n ({\bf t}, \bar {\bf t}, z)
\psi_{n'}^{\dag}({\bf t}', \bar {\bf t}', z)=
\oint_{C_0}\! \frac{dz}{z}\, \bar \psi_n ({\bf t}, \bar {\bf t}, z)
\bar \psi_{n'}^{\dag}({\bf t}', \bar {\bf t}', z).
\eeq

\subsection{The two-point Baker-Akhiezer functions}

The two-point Baker-Akhiezer function 
$\Psi =\Psi (n, {\bf t}, \bar {\bf t}, P)$ is a function
on the curve $\Gamma$ of genus $g$ 
which is analytic everywhere except at the marked points $P_{\infty}, P_0$, where
it has essential singularities of the prescribed form and $g$ simple poles at some 
points $Q_1, \ldots , Q_g \in \Gamma$. As before, we denote the divisor of poles of $\Psi$ by 
${\cal D}=Q_1+ \ldots + Q_g$
and assume that it is non-special. 
The standard argument shows that the linear space of two-point Baker-Akhiezer functions is
one-dimensional, i.e., it is unique up to a common
constant multiplier.
This
multiplier can be fixed by the condition $\displaystyle{\lim_{P\to P_\infty}\Bigl (
k^{-n}e^{-\xi ({\bf t}, k)}\Psi (n,{\bf t}, \bar {\bf t}, P)\Bigr )=1}$, where $k^{-1}$ is
the local parameter in a neighborhood of the point $P_{\infty}$ ($k(P_{\infty})=\infty$).
In the vicinity of $P_{\infty}$
the Baker-Akhiezer function has the form 
\beq\label{toda6}
\Psi (n,{\bf t}, \bar {\bf t}, P)=k^n e^{\xi ({\bf t}, k)}
\Bigl (1+\xi_1 (n,{\bf t}, \bar {\bf t})k^{-1} +\xi_2(n,{\bf t}, 
\bar {\bf t}) k^{-2} +\ldots \Bigr ), \quad P\to P_{\infty}
\eeq
(we assume that only a finite
number of the times $t_j$, $\bar t_j$ are different from zero).
In the vicinity of the other marked point, $P_{0}$, with the local
parameter $k$ ($k(P_0)=0$), the Baker-Akhiezer function has the form
\beq\label{toda7}
\Psi (n,{\bf t}, \bar {\bf t}, P)=k^n e^{\xi (\bar {\bf t}, k^{-1})+
\varphi_n({\bf t}, \bar {\bf t})}
\Bigl (1+\bar \xi_1 (n,{\bf t}, \bar {\bf t})k +\bar \xi_2(n,{\bf t}, 
\bar {\bf t}) k^2 +\ldots \Bigr ), \quad P\to P_0.
\eeq

In this section we denote by $d\Omega$ the abelian differential of the third kind 
with simple poles at the points $P_0, P_{\infty}$ with residues $\pm 1$
and $g$ simple zeros at the points of the divisor ${\cal D}$. Such a differential is unique
up to a common constant multiplier. It has other $g$ zeros at some points
$Q_1^{\dag}, \ldots , Q_g^{\dag}$. Let 
${\cal D}^{\dag}=Q_1^{\dag}+ \ldots + Q_g^{\dag}$
be the divisor of these zeros, then we have the relation
\beq\label{toda8}
{\cal D}+{\cal D}^{\dag}={\cal K}+P_{\infty}+P_0
\eeq
which means that
\beq\label{toda9}
\vec A({\cal D})+\vec A({\cal D}^{\dag})+2\vec K -\vec A(P_{\infty})-\vec A(P_{0})=0
\eeq
in the Jacobian. Similarly to the one-point case, one can introduce the
dual Baker-Akhiezer function 
$\Psi^{\dag}$ by the requirement that it
has the divisor of poles ${\cal D}^{\dag}$ and 
in the vicinity of the points $P_{\infty}$, $P_0$ it 
behaves as
\beq\label{toda10}
\Psi^{\dag} (n,{\bf t}, \bar {\bf t}, P)=k^{-n} e^{-\xi ({\bf t}, k)}
\Bigl (1+\xi_1^{\dag} (n,{\bf t}, \bar {\bf t})k^{-1} +\xi_2^{\dag}(n,{\bf t}, 
\bar {\bf t}) k^{-2} +\ldots \Bigr ), \quad P\to P_{\infty},
\eeq\beq\label{toda11}
\Psi^{\dag} (n,{\bf t}, \bar {\bf t}, P)=k^{-n} e^{-\xi (\bar {\bf t}, k^{-1})-
\varphi_n({\bf t}, \bar {\bf t})}
\Bigl (1+\bar \xi_1^{\dag} (n,{\bf t}, \bar {\bf t})k +\bar \xi_2^{\dag}(n,{\bf t}, 
\bar {\bf t}) k^2 +\ldots \Bigr ), \quad P\to P_0.
\eeq

The differential $\Psi (n, {\bf t}, \bar {\bf t}, P)\Psi^{\dag} (n', {\bf t}', \bar {\bf t}', P)
d\Omega (P)$ is holomorphic 
everywhere on $\Gamma$ except the points $P_{\infty}$, $P_0$. 
Therefore, sum of the residues at these points is equal to zero and we have
\beq\label{toda12}
\Bigl (\oint_{C_{\infty}} -\oint_{C_0}\Bigr ) \Psi (n,{\bf t}, \bar {\bf t}, P)
\Psi^{\dag}(n', {\bf t}', \bar {\bf t}', P)d\Omega (P) 
=0
\eeq
for all $n, n'$, 
${\bf t}, {\bf t}'$, where $C_{\infty}$ ($C_0$) is a small contour around 
the point $P_{\infty}$ ($P_0$).

The explicit form of the two-point Baker-Akhiezer function is
\beq\label{toda13}
\begin{array}{l}
\displaystyle{
\Psi (n,{\bf t}, \bar {\bf t}, P)= \exp \Bigl (n\Omega_0(P)-n\Omega_{00}+
\sum_{j\geq 1}t_j \Omega_j (P)+\sum_{j\geq 1}\bar t_j (\bar \Omega_j (P)
-\bar \Omega_j (P_{\infty})\Bigr )}
\\ \\
\phantom{aaaaaaa}\displaystyle{\times
\frac{\Theta \Bigl (\vec A(P)\! +\! n\vec U_0\! +\! \sum\limits_{j\geq 1}\vec U_j t_j 
\! +\! \sum\limits_{j\geq 1}\vec V_j \bar t_j\!-\!
\vec A({\cal D}) \!-\! \vec K\Bigr )\Theta \Bigl (\vec A({\cal D}) \! +\!
\vec K\! -\! \vec A(P_{\infty})\Bigr )}{\Theta \Bigl (\vec A(P)\! -\! 
\vec A({\cal D}) \! -\! \vec K\Bigr )\Theta \Bigl (
n\vec U_0\! +\! \sum\limits_{j\geq 1}\vec U_j t_j 
\! +\! \sum\limits_{j\geq 1}\vec V_j \bar t_j
\! -\! \vec A({\cal D}) \! -\!
\vec K\! +\! \vec A(P_{\infty})\Bigr ) }},
\end{array}
\eeq
where we use the notation introduced in section \ref{section:riemann}.
It is assumed that
the path of integration in the Abel map and $\Omega_j$, $\bar \Omega_j$ is the same. 
Tending $P\to P_0$ in (\ref{toda13}) 
and comparing with (\ref{toda7}), we obtain:
\beq\label{toda14}
\begin{array}{l}
\displaystyle{
e^{\varphi_n({\bf t}, \bar {\bf t})}=\exp \Bigl (n(\bar \Omega_{00}-
\Omega_{00})+
\sum_{j\geq 1}(t_j\Omega_j(P_0)-
\bar t_j \bar \Omega_j(P_{\infty})\Bigr )}
\\ \\
\phantom{a}\displaystyle{\times \,
\frac{\Theta \Bigl (\vec A ({\cal D})\! +\!
\vec K\! -\! \vec A(P_{\infty})\Bigr )\,
\Theta \Bigl ((n+1)\vec U_0 \! +\! \sum\limits_{j\geq 1}\vec U_j t_j 
\! +\! \sum\limits_{j\geq 1}\vec V_j \bar t_j\!-\!
\vec A({\cal D}) \!-\! \vec K \! +\! \vec A(P_{\infty})\Bigr )}{\Theta \Bigl (
\vec A ({\cal D})\! +\!
\vec K\! -\! \vec A(P_{0})\Bigr )\, 
\Theta \Bigl (n\vec U_0 \! +\! \sum\limits_{j\geq 1}\vec U_j t_j 
\! +\! \sum\limits_{j\geq 1}\vec V_j \bar t_j\!-\!
\vec A({\cal D}) \!-\! \vec K \! +\! \vec A(P_{\infty})\Bigr )}}.
\end{array}
\eeq
The explicit form of the dual Baker-Akhiezer function is
\beq\label{toda15}
\begin{array}{l}
\displaystyle{
\Psi^{\dag} (n,{\bf t}, \bar {\bf t}, P)= \exp \Bigl (-n\Omega_0(P)
+n\Omega_{00}-
\sum_{j\geq 1}t_j \Omega_j (P)-\sum_{j\geq 1}\bar t_j (\bar \Omega_j (P)
-\bar \Omega_j (P_{\infty})\Bigr )}
\\ \\
\phantom{aaaaaaaaaa}\displaystyle{\times
\frac{\Theta \Bigl (\vec A({\cal D}) \! +\!
\vec K\! -\! \vec A(P_{0})\Bigr )}{\Theta \Bigl (\vec A(P)\! +\! 
\vec A({\cal D}) \! +\! \vec K\! -\! \vec A(P_{\infty})\! -\! \vec A(P_{0})\Bigr )}}
\\ \\
\displaystyle{\times \, 
\frac{\Theta \Bigl (-\vec A(P)\! +\! (n+1)\vec U_0\! +\! \sum\limits_{j\geq 1}\vec U_j t_j 
\! +\! \sum\limits_{j\geq 1}\vec V_j \bar t_j\!-\!
\vec A({\cal D}) \!-\! \vec K\! +\! 2\vec A(P_{\infty})\Bigr )}{\Theta \Bigl (
(n+1)\vec U_0\! +\! \sum\limits_{j\geq 1}\vec U_j t_j 
\! +\! \sum\limits_{j\geq 1}\vec V_j \bar t_j
\! -\! \vec A({\cal D}) \! -\!
\vec K\! +\! \vec A(P_{\infty})\Bigr ) }}.
\end{array}
\eeq

The standard argument shows that the Baker-Akhiezer function satisfies the linear equation
\beq\label{lin3}
\p_{t_1}\Psi (n)=\Psi (n+1)+v_n \Psi (n), 
\eeq
where
\beq\label{lin4}
v_n ({\bf t}, \bar {\bf t})=\p_{t_1}\varphi_n ({\bf t}, \bar {\bf t}).
\eeq

\subsection{The tau-function}

The tau-function for quasi-periodic solutions of the 2D Toda lattice is given by
\beq\label{toda16}
\tau_n({\bf t}, \bar {\bf t})=\exp \Bigl (-
Q(n, {\bf t}, \bar {\bf t})\Bigr )\, \Theta \Bigl (
n\vec U_0 \! +\! \sum\limits_{j\geq 1}\vec U_j t_j 
\! +\! \sum\limits_{j\geq 1}\vec V_j \bar t_j +\vec Z\Bigr ),
\eeq
where the constant 
vector $\vec Z$ is defined in (\ref{tau1a}) and $Q(n, {\bf t}, \bar {\bf t})$
is the following quadratic form of the times $n, {\bf t}, \bar {\bf t}$:
\beq\label{toda17}
\begin{array}{l}
\displaystyle{
Q(n, {\bf t}, \bar {\bf t})=\frac{1}{2}\sum_{i,j}\Omega_{ij}t_it_j+
\frac{1}{2}\sum_{i,j}\bar \Omega_{ij}\bar t_i\bar t_j+\sum_{i,j}
\omega_{ij}t_i\bar t_j}
\\ \\
\phantom{aaaaaaaaaaaaaaaaaaaaa}\displaystyle{+
n\sum_j \Bigl (\Omega_{0j}t_j +\bar \Omega_{0j}\bar t_j\Bigr )
+\frac{1}{2}(\Omega_{00}-\bar \Omega_{00})n^2.}
\end{array}
\eeq
Here we use the notation introduced in section \ref{section:riemann}.

In the simplest possible case $\Gamma$ is a genus $0$ Riemann surface (the Riemann sphere).
In this case the theta-function is absent, 
$\Omega_j =k^j$, $\bar \Omega_j=k^{-j}$, $\Omega_0=\log k$ and so
$\Omega_{ij}=\bar \Omega_{ij}=0$, $\omega_{ij}=\delta_{ij}$. Therefore, the tau-function
in this case is the trivial one given by (\ref{simplest}). 

Using the relations presented in section \ref{section:riemann}, one can verify by 
straightforward calculations that
\beq\label{toda18}
\Psi (n, {\bf t}, \bar {\bf t}, P)=\left \{
\begin{array}{l}
\displaystyle{
k^n e^{\xi ({\bf t}, k)}\frac{\tau_0(0,0)}{\tau_0 (-[k^{-1}], 0)}\,
\frac{\tau_n ({\bf t}-[k^{-1}], \bar {\bf t})}{\tau_n ({\bf t}, \bar {\bf t})},
\quad P\to P_{\infty} ,}
\\ \\
\displaystyle{
k^n e^{\xi (\bar {\bf t}, k^{-1})+\omega_0}
\frac{\tau_0(0,0)}{\tau_1 (0, -[k])}\,
\frac{\tau_{n+1} ({\bf t}, \bar {\bf t}-[k])}{\tau_n ({\bf t}, \bar {\bf t})},
\, \quad P\to P_{0} ,}
\end{array}
\right.
\eeq

\beq\label{toda19}
\Psi^{\dag} (n, {\bf t}, \bar {\bf t}, P)=\left \{
\begin{array}{l}
\displaystyle{
k^{-n} e^{-\xi ({\bf t}, k)}\frac{\tau_1(0,0)}{\tau_1 ([k^{-1}], 0)}\,
\frac{\tau_{n+1} ({\bf t}+[k^{-1}], \bar {\bf t})}{\tau_{n+1} ({\bf t}, \bar {\bf t})},
\quad P\to P_{\infty} ,}
\\ \\
\displaystyle{
k^{-n} e^{-\xi (\bar {\bf t}, k^{-1})-\omega_0}
\frac{\tau_1(0,0)}{\tau_0 (0, [k])}\,
\frac{\tau_{n} ({\bf t}, \bar {\bf t}+[k])}{\tau_{n+1}({\bf t}, \bar {\bf t})},
\;\;\;\, \quad P\to P_{0} .}
\end{array}
\right.
\eeq
Here $\omega_0=\frac{1}{2}(\Omega_{00}-\bar \Omega_{00})$.
Recalling equations (\ref{toda4})--(\ref{toda4c}), we see that near the points 
$P_{\infty}$, $P_0$ the Baker-Akhiezer functions essentially coincide (up to
normalization) with the wave functions:
\beq\label{toda20}
\Psi (n, {\bf t}, \bar {\bf t}, P)=\left \{
\begin{array}{l}
\displaystyle{
\frac{\tau_0(0,0)}{\tau_0 (-[k^{-1}], 0)}\,
\psi_n ({\bf t}, \bar {\bf t},k), \quad k\to \infty , }
\\ \\
\displaystyle{\frac{\tau_0(0,0)e^{\omega_0}}{\tau_1 (0, -[k])}\,
\bar \psi_n ({\bf t}, \bar {\bf t},k), \phantom{a} \; \quad k\to 0,}
\end{array}
\right.
\eeq

\beq\label{toda21}
\Psi^{\dag} (n, {\bf t}, \bar {\bf t}, P)=\left \{
\begin{array}{l}
\displaystyle{
\frac{\tau_1(0,0)}{\tau_1 ([k^{-1}], 0)}\,
\psi^{\dag}_n ({\bf t}, \bar {\bf t},k), \quad k\to \infty , }
\\ \\
\displaystyle{\frac{\tau_1(0,0)e^{-\omega_0}}{\tau_0 (0, [k])}\,
\bar \psi^{\dag}_n ({\bf t}, \bar {\bf t},k), \phantom{a}\; \quad k\to 0,}
\end{array}
\right.
\eeq

It remains to identify the bilinear relations (\ref{toda5}) and (\ref{toda12}). 
Equations (\ref{toda20}), (\ref{toda21}) state that they coincide if the following
identities hold:
\beq\label{toda23}
d\Omega (P)=\frac{\tau_0 (-[k^{-1}], 0)\tau_1 ([k^{-1}], 0)}{\tau_0(0,0)\tau_1(0,0)}\,
\frac{dk}{k}, \quad k\to \infty ,
\eeq
\beq\label{toda23a}
d\Omega (P)=-\frac{\tau_1 (0, -[k])\tau_0 (0, [k])}{\tau_0(0,0)\tau_1(0,0)}\,
\frac{dk}{k}, \quad k\to 0.
\eeq
Let us prove the first one. We have:
$$
\mbox{The r.h.s. of (\ref{toda23})}\; 
$$
$$
=\exp \Bigl (-\!\sum_{i,j}\Omega_{ij}\frac{k^{-i-j}}{ij}-\! 
\sum_j \Omega_{0j}\frac{k^{-j}}{j}\Bigr )
\frac{\Theta \Bigl (\vec A(P)\! -\! \vec A(P_{\infty})\! +\! \vec Z\Bigr )
\Theta \Bigl (\vec A(P) \! -\! \vec A(P_{\infty})\! -\! 
\vec U_0\! -\! \vec Z\Bigr )}{\Theta \Bigl (\vec Z\Bigr )\,
\Theta \Bigl (\vec U_0 +\vec Z\Bigr )}\frac{dk}{k}
$$
$$
= C \frac{\Theta \Bigl (\vec A(P)\! -\! \vec A({\cal D})\! -\! \vec K\Bigr )
\Theta \Bigl (\vec A(P)\! -\! \vec A({\cal D}^{\dag})\! -\! \vec K\Bigr )}{\Theta_{*}^2
\Bigl (\vec A(P)-\vec A(P_{\infty})\Bigr )
\Theta \Bigl (\vec Z\Bigr )\,
\Theta \Bigl (\vec U_0 +\vec Z\Bigr )} \, 
\exp \Bigl (-\Omega_0(P)+\Omega_{00}\Bigr )\, d\zeta 
$$
$$
=\frac{\Theta \Bigl (\vec A(P)\! -\! \vec A({\cal D})\! -\! \vec K\Bigr )
\Theta \Bigl (\vec A(P)\! -\! 
\vec A({\cal D}^{\dag})\! -\! \vec K\Bigr )\Theta_{*}
\Bigl (\vec A(P_0)-\vec A(P_{\infty})\Bigr )
\, d\zeta}{\Theta \Bigl (\vec Z\Bigr )\,
\Theta \Bigl (\vec U_0 +\vec Z\Bigr )\Theta_{*}
\Bigl (\vec A(P)\! -\! \vec A(P_{\infty})\Bigr )
\Theta_{*}
\Bigl (\vec A(P)\! -\! \vec A(P_{0})\Bigr )},
$$
where $d\zeta$ is the differential (\ref{qp86}) and when passing from the second
to the third line we have used (\ref{qp85}) and 
(\ref{toda9}). One can see that the r.h.s. is a well-defined meromorphic differential
on $\Gamma$ with the only simple poles at the points $P_{\infty}$, $P_0$ and simple zeros
at the divisors ${\cal D}$, ${\cal D}^{\dag}$. Therefore, it must be proportional
to $d\Omega$. The identity (\ref{toda23a}) can be proved in a similar way.

\section{The BKP hierarchy}

\subsection{Bilinear relations for the BKP hierarchy}

The BKP hierarchy was introduced in \cite{DJKM81,DJKM82}.
The set of independent variables in the BKP hierarchy is the infinite set of
``times'' $t_j$ numbered by odd indices: ${\bf t}_{\rm o}=\{t_1, t_3, t_5, \ldots \, \}$. 
The universal dependent variable is the tau-function $\tau ({\bf t}_{\rm o})$.
All equations of the hierarchy are encoded in the bilinear relation
\beq\label{bkp1}
\oint_{C_{\infty}}e^{\xi ({\bf t}_{\rm o}-{\bf t}'_{\rm o}, z)}
\tau ({\bf t}_{\rm o}-2[z^{-1}]_{\rm o})\tau ({\bf t}'_{\rm o}+2[z^{-1}]_{\rm o})\,
\frac{ dz}{2\pi i z}=\tau ({\bf t}_{\rm o})\tau ({\bf t}'_{\rm o})
\eeq
valid for arbitrary ${\bf t}_{\rm o}$, ${\bf t}'_{\rm o}$, where
$$
\begin{array}{c}
{\bf t}_{\rm o}\pm 2[z^{-1}]_{\rm o}=\Bigl \{ t_1 \pm 2z^{-1}, t_3\pm \frac{2}{3}z^{-3},
t_5\pm \frac{2}{5}z^{-5}, \ldots \Bigr \}
\end{array}
$$
and 
\beq\label{bkp2}
\xi ({\bf t}_{\rm o}, z)=\! \sum_{j\geq 1, \,\, {\rm odd}} t_j z^j.
\eeq
The contour $C_{\infty}$ is a big circle encircling $\infty$. 
The simplest (trivial) solution of (\ref{bkp1}) is $\tau ({\bf t}_{\rm o})=1$.

Set 
${\bf t'}_{\rm o}={\bf t}_{\rm o}-2[a^{-1}]_{\rm o}-
2[b^{-1}]_{\rm o}-2[c^{-1}]_{\rm o}$ in (\ref{bkp1}), then
$$
e^{\xi ({\bf t}_{\rm o}-{\bf t}_{\rm o}', z)}=\frac{(a+z)(b+z)(c+z)}{(a-z)(b-z)(c-z)}
$$
and the residue calculus gives the following equation:
\beq\label{bkp3}
\begin{array}{c}
(a+b)(a+c)(b-c)\tau^{[a]}\tau^{[bc]}
+ (b+a)(b+c)(c-a)\tau^{[b]}\tau^{[ac]}
\\ \\
+ (c+a)(c+b)(a-b)\tau^{[c]}\tau^{[ab]}
+(a-b)(b-c)(c-a)\tau \tau^{[abc]}=0,
\end{array}
\eeq
where
$
\tau^{[a]}:=\tau ({\bf t_{\rm o}}+2[a^{-1}]_{\rm o})$,
$\tau^{[ab]}:=\tau ({\bf t}_{\rm o}+2[a^{-1}]_{\rm o}+2[b^{-1}]_{\rm o})$,
$\tau^{[abc]}:=\tau ({\bf t}_{\rm o}+2[a^{-1}]_{\rm o}+2[b^{-1}]_{\rm o}+2[c^{-1}]_{\rm o})$. 
For the first time it has appeared in \cite{Miwa}.
Taking the limit $c\to \infty$, we get the equation
\beq\label{bkp4}
\tau \tau^{[ab]}\left (1+\frac{1}{a+b}\, \p_{t_1}\log \frac{\tau}{\tau^{[ab]}}\right )
=\tau^{[a]}\tau^{[b]}
\left (1+\frac{1}{a-b}\, \p_{t_1}\log \frac{\tau^{[b]}}{\tau^{[a]}}\right ).
\eeq
The differential equations of the hierarchy are obtained by expanding it in 
inverse powers of $a,b$.

The wave function $\psi ({\bf t}_{\rm o}, z)$ is connected with the tau-function 
by the formula
\beq\label{bkp5}
\psi ({\bf t}_{\rm o}, z)=
\psi^{\rm BKP}({\bf t}_{\rm o}, z)
=e^{\xi ({\bf t}_{\rm o}, z)}\frac{\tau ({\bf t}_{\rm o}-
2[z^{-1}]_{\rm o})}{\tau ({\bf t}_{\rm o})}.
\eeq
In terms of the wave function, the bilinear relation acquires the form
\beq\label{bkp6}
\oint_{C_{\infty}}\psi ({\bf t}_{\rm o}, z)\, \psi ({\bf t}'_{\rm o}, -z)\, 
\frac{dz}{2\pi i z} =1.
\eeq

The KP and BKP hierarchies are closely related with each other. Their interrelations
were discussed in \cite{DJKM82a,DJKM83}, see also the recent review \cite{Z21}.
The latter can be regarded as a subhierarchy of the former, with all times with even 
indices being fixed to $0$ and some strong restrictions in the space of solutions. 
Let ${\bf t}_{\bullet}$ be the set of times
${\bf t}_{\bullet}=\{t_1, 0, t_3, 0, t_5, 0, \ldots \}$ and let $\tau^{\rm KP}({\bf t})$
be the KP tau-function. The following constraint distinguishes those KP tau-functions
that give rise to solutions of the BKP hierarchy:
\beq\label{bkp7}
\Bigl ( \tau^{\rm KP}({\bf t}_{\bullet}-[z^{-1}])\Bigr )^2=
\tau^{\rm KP}({\bf t}_{\bullet})\tau^{\rm KP}({\bf t}_{\bullet}-2[z^{-1}]_{\rm o}).
\eeq
Moreover, the BKP tau-function is the square root of the KP tau-function:
\beq\label{bkp8}
\tau ({\bf t}_{\rm o})=\sqrt{\tau^{\rm KP}({\bf t}_{\bullet})}.
\eeq
Taking this into account, one can rewrite (\ref{bkp7}) in the form
\beq\label{bkp9}
\tau^{\rm KP}({\bf t}_{\bullet}-[z^{-1}])=\tau ({\bf t}_{\rm o})\,
\tau ({\bf t}_{\rm o}-2[z^{-1}]_{\rm o}).
\eeq
Comparing (\ref{kp5}) and (\ref{bkp5}), we then conclude that
\beq\label{bkp10}
\psi^{\rm KP}({\bf t}_{\bullet}, z)=\psi^{\rm BKP} ({\bf t}_{\rm o}, z),
\eeq
so the wave functions in the two hierarchies are essentially the same.

\subsection{The Baker-Akhiezer function}

The quasi-periodic solutions of the BKP hierarchy can be constructed along the same
lines as those of the KP hierarchy. However, while in the latter case one could start
from an arbitrary genus $g$ algebraic curve and arbitrary effective divisor of degree $g$
on it, in the BKP case the algebraic-geometrical data underlying the solution are not
arbitrary. Namely, the algebraic curve must admit a holomorphic involution $\iota$
with two fixed points one of which is $P_{\infty}$ (and the other one is $P_0$).
From the Riemann-Hurwitz formula it then follows that $g$ is even: $g=2g_0$. 
The divisor is also subject to some restriction (see (\ref{invb1}) below). 

The Baker-Akhiezer function 
is defined in the same way as in the beginning of section \ref{section:baker}.
As we have seen, it essentially coincides with 
the wave function. Under the conditions outlined above,
it can be explicitly constructed either in 
terms of the Riemann theta-functions, as in section \ref{section:baker}
(the left hand side of (\ref{bkp10}), or in terms of the Prym theta-functions
(the right hand side of (\ref{bkp10}). The identity (\ref{bkp10}) states that these two
ways are equivalent and lead to the same result.

\paragraph{Admissible divisors.}
We call the effective divisor ${\cal D}=Q_1+\ldots +Q_g$ {\it admissible} if $2g$ points
$Q_i$, $\iota Q_i$ are zeros of a meromorphic differential $d\Omega$ which has simple
poles at the fixed points of the involution $P_{0}$ (with residue
$+1$) and $P_\infty$ (with residue $-1$). In other words, admissible divisors are
subject to the condition
\beq\label{invb1}
{\cal D}+\iota {\cal D}={\cal K}+P_0 +P_{\infty},
\eeq
where ${\cal K}$ is the canonical class. Taking the Abel map (we recall that the initial
point of the Abel map is $P_0$), we have
$$
\vec A({\cal D})+\vec A(\iota {\cal D})+2\vec K =\vec A(P_{\infty}),
$$
or
\beq\label{invb2}
\begin{array}{c}
\iota \Bigl (\vec A({\cal D})-\frac{1}{2}\, \vec A(P_{\infty})+\vec K\Bigr )=
-\Bigl (\vec A({\cal D})-\frac{1}{2}\, \vec A(P_{\infty})+\vec K\Bigr ),
\end{array}
\eeq
whence we conclude that 
\beq\label{invb3}
\begin{array}{c}
\vec z=-
\vec A({\cal D})-\vec K +\frac{1}{2} \vec A(P_{\infty})
\end{array}
\eeq
belongs to the Prym variety (cf. (\ref{tau1a})). In some formulas 
below we treat $\vec z$ as a vector 
belonging to $\CC^{g_0}$ (via the inverse map $\sigma^{-1}$). 

\paragraph{Baker-Akhiezer function in terms of Prym theta-functions.}
The explicit form of the Baker-Akhiezer function in terms of the Prym theta-functions 
and Abel-Prym map is
\beq\label{invb4}
\Psi ({\bf t}_{\rm o}, P)=
\exp \Bigl (\sum_{j\geq 1, \, {\rm odd}}t_j \Omega_j (P)\Bigr )
\frac{\Theta_{\rm Pr}\Bigl (\vec A^{\rm Pr}(P)+\!\!\sum\limits_{j\geq 1,\, {\rm odd}}
\vec U_j t_j+\vec z\Bigr )\Theta_{\rm Pr}
\Bigl (\vec z)\Bigr )}{\Theta_{\rm Pr}\Bigl (
\vec A^{\rm Pr}(P)+\vec z\Bigr )
\Theta_{\rm Pr}\Bigl (\sum\limits_{j\geq 1,\, {\rm odd}}
\vec U_j t_j+\vec z\Bigr )}.
\eeq
This function has poles at the divisor ${\cal D}$ and the essential singularity
at the point $P_{\infty}$ with the expansion of the same form 
(\ref{ba1}). The differential $\Psi ({\bf t}_{\rm o}, P)
\Psi ({\bf t}_{\rm o}, \iota P)d\Omega$ is holomorphic everywhere except the points
$P_0$, $P_{\infty}$, where it has simple poles. 
Noting that $\lim\limits_{P\to P_{\infty}}
\Psi ({\bf t}_{\rm o}, P)\Psi ({\bf t}_{\rm o}, \iota P)=1$
and using the residue theorem, we conclude that 
$\Psi({\bf t}_{\rm o}, P_0)=\pm 1$. 
Since $\Psi(0, P_0)=1$ and this function is continuous 
in ${\bf t}_{\rm o}$, we see that we should choose the plus sign:
\beq\label{invb7}
\Psi({\bf t}_{\rm o}, P_0)=1.
\eeq

Consider now the differential $\Psi ({\bf t}_{\rm o}, P)
\Psi ({\bf t}'_{\rm o}, \iota P)d\Omega$. This differential, too, is
holomorphic everywhere on $\Gamma$ except the points
$P_0$, $P_{\infty}$, where it has simple poles. Since $\Psi({\bf t}_{\rm o}, P_0)=
\Psi({\bf t}'_{\rm o}, P_0)=1$, the 
residue at $P_0$ is equal to $1$, whence the residue at 
$P_{\infty}$ is equal to $-1$
and thus the bilinear relation
\beq\label{invb5}
\oint_{C_{\infty}}\! \Psi ({\bf t}_{\rm o}, P)
\Psi ({\bf t}'_{\rm o}, \iota P)d\Omega =1
\eeq
holds for all ${\bf t}_{\rm o}$, ${\bf t}'_{\rm o}$.

The standard argument based on uniqueness of 
the Baker-Akhiezer function shows that it satisfies the 
linear equation
\beq\label{invb6}
\p_{t_3}\Psi =\p_x^3\Psi +3u \p_x \Psi , \quad x=t_1, \quad u=\p_x \xi_1 .
\eeq

\paragraph{Baker-Akhiezer function in terms of Riemann theta-functions.}
Alternatively, one can give an explicit expression for the Baker-Akhiezer function
in terms of the Riemann theta-functions and Abel map according to (\ref{ba5}):
\beq\label{invb8}
\Psi ({\bf t}_{\rm o}, P)=
\exp \Bigl (\sum_{j\geq 1, \, {\rm odd}}t_j \Omega_j (P)\Bigr )
\frac{\Theta \Bigl (\vec A(P)-\vec A(P_{\infty})+\!\sum\limits_{j\geq 1,\, {\rm odd}}
\vec U_j t_j +\vec Z\Bigr )\Theta \Bigl (\vec Z\Bigr )}{\Theta
\Bigl (\vec A(P)-\vec A(P_{\infty})+\vec Z\Bigr )\Theta \Bigl (
\sum\limits_{j\geq 1,\, {\rm odd}} \vec U_j t_j+\vec Z\Bigr )},
\eeq
\beq\label{invb9}
\vec Z =\vec z +\frac{1}{2}\vec A(P_{\infty}).
\eeq
The equivalence of (\ref{invb4}) and (\ref{invb8}) is based on the following 
remarkable identity:
\beq\label{invb10}
\Theta ^2 \Bigl (\vec A(P)-\vec A(P_{\infty })
+\vec Z_{{\bf t}_{\rm o}}\Bigr )=C(P)\Theta \Bigl (\vec A(P)-\vec A(\iota P)
+\vec Z_{{\bf t}_{\rm o}}\Bigr )\Theta \Bigl (\vec Z_{{\bf t}_{\rm o}}\Bigr )
\eeq
valid for any vector $\vec Z_{{\bf t}_{\rm o}} =-\vec A({\cal D})-
\vec K +\vec A(P_{\infty })+\! \sum\limits_{j\geq 1,\, {\rm odd}}
\vec U_j t_j=\vec Z +\! \sum\limits_{j\geq 1,\, {\rm odd}}
\vec U_j t_j$, where
${\cal D}$ is an admissible divisor. Indeed, substituting (\ref{invb10}) into 
(\ref{invb8}) and using (\ref{inv8}), we get (\ref{invb4}). We note that 
identity (\ref{invb10}) is nothing else than identity (\ref{bkp7}) for the
KP tau-functions that give rise to solutions of the BKP hierarchy.

\paragraph{Proof of identity (\ref{invb10}).}
The proof of (\ref{invb10}) goes along the lines of the proof of a similar
identity given in \cite{KZ22}. This proof belongs to I.Krichever.
For each $Q\in \Gamma$, $Q\neq P_0, P_{\infty}$, 
let $d\Omega_Q(P)$ be the unique normalized
meromorphic differential of the third kind with simple 
poles at $Q$ and $\iota Q$ with
residues $+1$ and $-1$ respectively and let
$\Omega_Q(P)$ be the corresponding abelian integral
\beq\label{pr0}
\Omega_Q(P)=\int_{P_{\infty}}^Pd\Omega_Q .
\eeq

Consider the function $\Psi_Q({\bf t}_{\rm o}, P)$
defined by the formula
\beq\label{pr1}
\begin{array}{l}
\displaystyle{
\Psi_Q({\bf t}_{\rm o}, P)=
\exp \Bigl (\Omega_Q(P)+\!\sum_{j\geq 1, \, {\rm odd}}t_j \Omega_j (P)\Bigr )}
\\ \\
\displaystyle{\phantom{aaaaaaaaaaaaa}\times \,
\frac{\Theta \Bigl (\vec A(P)-\vec A(P_{\infty})+\vec A(Q)-\vec A(\iota Q)
+\!\sum\limits_{j\geq 1,\, {\rm odd}}
\vec U_j t_j +\vec Z\Bigr )\Theta \Bigl (\vec Z\Bigr )}{\Theta
\Bigl (\vec A(P)-\vec A(P_{\infty})+\vec Z\Bigr )\Theta \Bigl (
\sum\limits_{j\geq 1,\, {\rm odd}} 
\vec U_j t_j+\vec A(Q)-\vec A(\iota Q)+\vec Z\Bigr )}}.
\end{array}
\eeq
It is easy to check that $\Psi_Q({\bf t}_{\rm o}, P)$ is the Baker-Akhiezer function
corresponding to the non-effective divisor ${\cal D}+\iota Q -Q$. Indeed, the Riemann
bilinear identity gives the relation $\displaystyle{
\frac{1}{2\pi i}\oint_{b_{\alpha}}d\Omega_Q =
\vec A(Q)-\vec A(\iota Q)}$. Hence the right hand side of (\ref{pr1}) does not depend
on a choice of the integration path. In a neighborhood of $P_{\infty}$ the function
$\Psi_Q({\bf t}_{\rm o}, P)$ has the form
\beq\label{pr2}
\Psi_Q({\bf t}_{\rm o}, P)=e^{\xi ({\bf t}_{\rm o}, k)}
\Bigl (1+\xi_{Q,1} ({\bf t}_{\rm o})k^{-1} +\xi_{Q,2}({\bf t}_{\rm o}) k^{-2} +
\ldots \Bigr ), \quad P\to P_{\infty}.
\eeq
Finally, this function has poles at ${\cal D}$ 
and an additional pole at $\iota Q$,
where the factor $e^{\Omega_Q(P)}$ in (\ref{pr1}) 
has simple pole. Furthermore, 
this factor vanishes at $P=Q$, therefore $\Psi_Q({\bf t}_{\rm o}, Q)=0$. 

Let us consider the differential 
$\Psi_Q({\bf t}_{\rm o}, P)\Psi_Q({\bf t}_{\rm o}, \iota P)d\Omega$. 
By the same argument
as above, it is a meromorphic differential with no poles at ${\cal D}$ and $\iota {\cal D}$.
A possible pole at $\iota Q$ is canceled by the zero of $\Psi_Q$ at $Q$. Hence this 
differential has simple poles at the points $P_0$ and $P_{\infty}$ only. Then the same
arguments as in the proof of (\ref{invb7}) give the equation
\beq\label{pr3}
\Psi_Q^2({\bf t}_{\rm o}, P_0)=1 \; \Rightarrow \; 
\Psi_Q({\bf t}_{\rm o}, P_0)=\pm 1.
\eeq
It turns out that, in contrast to (\ref{invb7}), 
the correct sign here is $-1$:
\beq\label{pr3a}
\Psi_Q({\bf t}_{\rm o}, P_0)=-1.
\eeq
(This looks like a counter-intuitive fact!) 
The correct sign can be determined by the following
argument. The function $\Psi_Q({\bf t}_{\rm o}, P_0)$ 
is continuous in ${\bf t}_{\rm o}$
and $Q$ for $Q\neq P_{\infty}$. Therefore, in order to find the correct sign,
we set ${\bf t}_{\rm o}=0$ and choose $Q$ in a 
neighborhood of $P_{\infty}$ with
local parameter $k^{-1}$. By the definition of $\Omega_Q(P)$ we have
$$
e^{\Omega_Q(P)}=\frac{k(P)-k(Q)}{k(P)+k(Q)}\, \Bigl (1+ O(k^{-1}(P))\Bigr ).
$$
For any fixed positive $r\ll 1$ one can choose 
$\eta$ such that for $|k^{-1}(Q)|<\eta$
and $|k^{-1}(P)|=r$ the inequality
$
\Bigl |1+e^{\Omega_Q(P)}\Bigr |<M\eta
$
holds with a constant $M$ independent of $\eta$. 
Moreover, for any fixed path from a point
$P'$, $|k^{-1}(P')|=r$, to $P_0$ there exists a constant $M_1$ (which depends on the path
but does not depend on $\eta$) such that the inequality
$\displaystyle{\Bigl |\int_{P'}^{P_0}d\Omega_Q 
\Bigr |<M_1\eta}$ holds. It then follows
that for a fixed path from $P_{\infty}$ to $P_0$
\beq\label{pr4}
\lim_{Q\to P_{\infty}}e^{\Omega_Q(P_0)}=-1.
\eeq
In fact if we define $\displaystyle{\tilde 
\Omega_Q(P)=\int_{Q_0}^Pd\Omega_Q}$ with
some initial point $Q_0\to P_{\infty}$, 
then the result depends on the order of
the limits:
$$
\lim_{Q_0\to P_{\infty}}\lim_{Q\to P_{\infty}}e^{\tilde \Omega_Q(P_0)}=1
\quad
\mbox{but}
\quad
\lim_{Q\to P_{\infty}}\lim_{Q_0\to P_{\infty}}e^{\tilde \Omega_Q(P_0)}=-1,
$$
and our case is the latter 
limit\footnote{We can illustrate the somewhat counter-intuitive
conclusion (\ref{pr4}) by the case when $\Gamma$ is the Riemann sphere, then
$$
\exp \int_{P_{\infty}}^{P_0}d\Omega_Q =\exp \int_{\infty}^{k_0}\Bigl (
\frac{dz}{z-q}-\frac{dz}{z+q}\Bigr )=\frac{k_0-q}{k_0+q}, \quad q=k(Q),
$$
and this indeed tends to $-1$ as $q\to \infty$.}. 
The ratio of theta-functions in (\ref{pr1}) is a continuous function of $Q$ which
tends to $1$ as $Q\to P_{\infty}$ and thus equation (\ref{pr3a}) is proved. 

Now consider the differential $\Psi_Q({\bf t}_{\rm o},P)\Psi ({\bf t}_{\rm o},\iota P)d\Omega$.
It is a meromorphic differential with possible poles at $P_0$, $P_{\infty}$ and
$\iota Q$. From (\ref{invb7}), (\ref{pr3a}) it follows that the sum of residues 
at $P_0$, $P_{\infty}$ equals $-2$. Therefore,
\beq\label{pr5}
\res_{P=\iota Q}\Bigl (\Psi_Q({\bf t}_{\rm o},P)\Psi ({\bf t}_{\rm o},\iota 
P)d\Omega (P)\Bigr )
=2.
\eeq
From (\ref{invb7}) it follows that the differential
$\Psi_Q({\bf t}_{\rm o},P)\p_{t_1}\Psi ({\bf t}_{\rm o},\iota P)d\Omega$
is holomorphic at $P_0$. Then the residue argument implies that
\beq\label{pr6}
\res_{P=\iota Q}\Bigl (\Psi_Q({\bf t}_{\rm o},P)
\p_{t_1}\Psi ({\bf t}_{\rm o},\iota P)d\Omega (P)\Bigr )=
-\res_{P=P_{\infty}}\Bigl (\Psi_Q({\bf t}_{\rm o},P)
\p_{t_1}\Psi ({\bf t}_{\rm o},\iota P)d\Omega (P)\Bigr ).
\eeq
The right hand side of (\ref{pr6}) can be expressed in terms of the first coefficients
$\xi_1$ and $\xi_{Q,1}$ of the expansions of $\Psi$ and $\Psi_Q$ around $P_{\infty}$:
\beq\label{pr7}
\res_{P=P_{\infty}}\Bigl (\Psi_Q({\bf t}_{\rm o},P)
\p_{t_1}\Psi ({\bf t}_{\rm o},\iota P)d\Omega (P)\Bigr )=
\xi_{Q,1}({\bf t}_{\rm o})-\xi_{1}({\bf t}_{\rm o}).
\eeq
From (\ref{pr5}), (\ref{pr6}) and (\ref{pr7}) it follows that
\beq\label{pr8}
\p_{t_1}\log \Psi ({\bf t}_{\rm o}, Q)=-\frac{1}{2}
\Bigl (\xi_{Q,1}({\bf t}_{\rm o})-\xi_{1}({\bf t}_{\rm o})\Bigr ).
\eeq
Equation (\ref{qp6}) implies that $\vec A(P)-\vec A(P_{\infty})=-\vec U_1 k^{-1}+
O(k^{-2})$. Using this relation, one can find the first coefficients of the expansion of
the ratios of theta-functions in the formulas for $\Psi$ and $\Psi_Q$ near $P_{\infty}$:
\beq\label{pr9}
\xi_{Q,1}({\bf t}_{\rm o})-\xi_{1}({\bf t}_{\rm o})=
c-\p_{t_1}\log \frac{\Theta \Bigl (\vec A(Q)-
\vec A(\iota Q)+\vec Z_{{\bf t}_{\rm o}}\Bigr )}{\Theta 
\Bigl (\vec Z_{{\bf t}_{\rm o}}\Bigr )},
\eeq
where
\beq\label{pr10}
\left. c=\frac{\p \Omega_Q(P)}{\p k^{-1}}\right |_{P=P_{\infty}}.
\eeq
The evaluation of formula (\ref{ba5}) at $P=Q$ yields
\beq\label{pr11}
\p_{t_1}\log \Psi ({\bf t}_{\rm o}, Q)=\p_{t_1}\log \frac{\Theta \Bigl (\vec A(Q)
+\vec Z_{{\bf t}_{\rm o}}\Bigr )}{\Theta 
\Bigl (\vec Z_{{\bf t}_{\rm o}}\Bigr )} +\Omega_1(Q).
\eeq
From the Riemann bilinear identity for the differentials $d\Omega_1$ and $d\Omega_Q$
it follows that
$$
c=\Omega_1(\iota Q)-\Omega_1(Q)=-2\Omega_1(Q).
$$
Then from (\ref{pr8}), (\ref{pr9}) and (\ref{pr11}) equation (\ref{invb10}) follows.

\subsection{The tau-function}

The tau-function of the BKP hierarchy is given by
\beq\label{taub1}
\tau ({\bf t}_{\rm o})=\exp \Bigl (-\frac{1}{4}\sum_{i,j \,\, {\rm odd}}
\Omega_{ij}t_it_j\Bigr )
\Theta_{\rm Pr} \Bigl (\sum\limits_{j\geq 1,\, {\rm odd}}
\vec U_j t_j+\vec z\Bigr ).
\eeq
Equation (\ref{inv8}) implies that it is indeed 
a square root of the KP tau-function (\ref{tau1}) at $t_{2j}=0$. 

It is instructive to compare the BKP tau-function with the mKP tau-function
$\tau_n^{\rm mKP}({\bf t})$ with even times put equal to zero. 
Using the connection between the mKP and Toda lattice hierarchies
explained in section \ref{section:todabilinear} and formula (\ref{toda16}) for the Toda
tau-function, we can write the algebraic-geometrical mKP tau-function associated with
curves with involution:
\beq\label{taub1a}
\tau_n^{\rm mKP}({\bf t})\Bigr |_{t_{2j}=0}
=\exp \Bigl (-\frac{1}{2} \sum_{i,j \,\, {\rm odd}}
\Omega_{ij}t_it_j\Bigr )\Theta \Bigl (
n\vec U_0+\!\! \sum\limits_{j\geq 1,\, {\rm odd}}
\vec U_j t_j+\vec Z\Bigr ).
\eeq
Here $U_0=\vec A(P_0)-\vec A(P_{\infty })=-\vec A(P_{\infty })$. From (\ref{inv8}) 
we see that
\beq\label{taub1b}
\tau_0^{\rm mKP}({\bf t})\Bigr |_{t_{2j}=0}=\tau_1^{\rm mKP}({\bf t})\Bigr |_{t_{2j}=0}=
\tau^2 ({\bf t}_{\rm o})
\eeq
and, furthermore,
\beq\label{taub1c}
\tau_n^{\rm mKP}({\bf t})\Bigr |_{t_{2j}=0}=\tau_{1-n}^{\rm mKP}({\bf t})\Bigr |_{t_{2j}=0},
\quad n\geq 0.
\eeq
The constraint (\ref{taub1c}) is specific for the B-version of the Toda (or mKP) hierarchy
discussed in \cite{UT84}. 

Similarly to the KP case, using (\ref{inv6}), we can see that
\beq\label{taub2}
\Psi ({\bf t}_{\rm o}, P)=e^{\xi ({\bf t}_{\rm o}, k)}
\frac{\tau (0)\, \tau ({\bf t}_{\rm o}-2[k^{-1}]_{\rm o})}{\tau ({\bf t}_{\rm o})
\tau (-2[k^{-1}]_{\rm o})}
=\frac{\tau (0)}{\tau (-2[k^{-1}]_{\rm o})}\, \psi ({\bf t}_{\rm o}, k),
\eeq
where the BKP wave function $\psi ({\bf t}_{\rm o}, k)$ is given by (\ref{bkp5}). 

Finally, we are going to prove that the bilinear relation (\ref{bkp6}) for the wave
function is the same as the bilinear relation (\ref{invb5}) for the Baker-Akhiezer function.
Their equivalence is based on the following identity:
\beq\label{taub3}
d\Omega (P)=\frac{\tau (-2[k^{-1}]_{\rm o})\tau (2[k^{-1}]_{\rm o})}{\tau^2 (0)} 
\, \frac{dk}{k}
\eeq
or
\beq\label{taub3a}
d\Omega (P)=\exp \Bigl (-2\! \sum_{i,j \, {\rm odd}}\Omega_{ij}\frac{k^{-i-j}}{ij}\Bigr )
\frac{\Theta_{\rm Pr}(\vec A^{\rm Pr}(P)+\vec z)
\Theta_{\rm Pr}(\vec A^{\rm Pr}(P)-\vec z)}{\Theta_{\rm Pr}^2(\vec z)}
\, \frac{dk}{k}.
\eeq
Using the relation (\ref{inv8}) we can
write this in terms of the Riemann theta-functions:
\beq\label{taub4}
\begin{array}{l}
\displaystyle{
d\Omega (P)=\exp \Bigl (-2\! \sum_{i,j \, {\rm odd}}\Omega_{ij}\frac{k^{-i-j}}{ij}\Bigr )
\, \frac{dk}{k}}
\\ \\
\displaystyle{\phantom{aaaaaaa}\times \left (
\frac{\Theta \Bigl (\vec A(P)\! -\! \vec A(\iota P)\! -\! 
\vec A({\cal D}) \! -\! \vec K \! +\! \vec A(P_{\infty })
\Bigr ) 
\Theta \Bigl (\vec A(P)\! -\! \vec A(\iota P)\! -\! \vec A({\cal D}) \! -\! 
\vec K \Bigr )}{\Theta^2
\Bigl (\vec Z\Bigr )}\right )^{1/2}} .
\end{array}
\eeq
Now, taking into account (\ref{an3}), we represent this identity in the form
\beq\label{taub4a}
\begin{array}{l}
\displaystyle{
d\Omega (P)=-\frac{2\sqrt{d\zeta (P)d\zeta (\iota P)}}{\Theta_{*}(\vec A(P)-\vec A(\iota P))}}
\\ \\
\displaystyle{\phantom{aaaaaaa}\times \left (
\frac{\Theta \Bigl (\vec A(P)\! -\! \vec A(\iota P)\! -\! 
\vec A({\cal D}) \! -\! \vec K \! +\! \vec A(P_{\infty })
\Bigr ) 
\Theta \Bigl (\vec A(P)\! -\! \vec A(\iota P)\! -\! \vec A({\cal D}) \! -\! 
\vec K \Bigr )}{\Theta^2
\Bigl (\vec Z\Bigr )}\right )^{1/2}} .
\end{array}
\eeq
One can check that the right hand side is a well-defined differential with zeros
at ${\cal D}$, $\iota {\cal D}$ and two simple poles at $P_0$, $P_{\infty}$ with 
residues $\pm 1$. Therefore, it indeed coincides with $d\Omega$.

\section*{Acknowledgments}

\addcontentsline{toc}{section}{Acknowledgments}

The research has been supported in part 
within the framework of the
HSE University Basic Research Program.

\end{document}